\documentclass[aps,prb,twocolumn,10pt,superscriptaddress,showpacs]{revtex4-2}
\usepackage{amsmath,amssymb,graphics,epsfig,epstopdf,color,verbatim,ulem,braket,tabularx,multirow,array,diagbox,colortbl,hhline}
\usepackage[colorlinks,linkcolor=blue,citecolor=blue,urlcolor=blue]{hyperref}
\usepackage{physics}
\usepackage{bm}
\usepackage{blindtext}
\usepackage{booktabs}
\usepackage{csquotes}
\usepackage{slashed}
\usepackage{siunitx}
\graphicspath{{Figures/}}

\begin{document}

\title{Pairing order and condensate fraction across the BCS-BEC crossover 
\\
in the two-dimensional attractive Hubbard model}

\author{Hao Du}
\thanks{These authors contributed equally to this work.}
\affiliation{Institute of Modern Physics, Northwest University, Xi'an 710127, China}

\author{Qian Sun}
\thanks{These authors contributed equally to this work.}
\affiliation{School of Physics, Northwest University, Xi'an 710127, China}

\author{Zhuotao Xie}
\affiliation{Institute of Modern Physics, Northwest University, Xi'an 710127, China}
\author{Yuan-Yao He}
\email{heyuanyao@nwu.edu.cn}
\affiliation{Institute of Modern Physics, Northwest University, Xi'an 710127, China}
\affiliation{Shaanxi Key Laboratory for Theoretical Physics Frontiers, Xi'an 710127, China}
\affiliation{Fundamental Discipline Research Center for Quantum Science and Technology of Shaanxi Province, Xi'an 710127, China}
\affiliation{Hefei National Laboratory, Hefei 230088, China}

\begin{abstract}
We report a systematic study of the superconducting ground state in the two-dimensional attractive Hubbard model, employing the {\it numerically exact} auxiliary-field quantum Monte Carlo method. We focus primarily on the pairing order and condensate fraction across the BCS-BEC crossover. Our numerical calculations cover a broad and physically relevant parameter regime, with the interaction strength $U/t=2\sim 12$ (where $t$ is the nearest-neighbor hopping amplitude) and fermion density $n=0.25\sim 1.0$. We systematically characterize the dependence of the pairing order and condensate fraction on both the interaction strength and fermion density, and perform finite-size extrapolations to obtain their thermodynamic-limit values. As complementary quantities, we also provide high-precision numerical results for the double occupancy and total energy. Furthermore, we investigate the effects of next-nearest-neighbor hopping on the superconducting properties of the model. Our results offer a comprehensive set of unbiased numerical data for the attractive Hubbard model, establishing quantitative benchmarks for understanding superconductivity across the BCS-BEC crossover and for validating future theoretical and numerical approaches.
\end{abstract}

\date{\today}

\maketitle

\section{Introduction}
\label{sec:intro}

The BCS-BEC crossover~\cite{Stefano2008,Leggett2012,Mohit2014,Giancarlo2018,Chen2005,Chen2024} provides a paradigmatic framework for understanding the evolution of superconductivity and fermionic superfluidity from weakly bound Cooper pairs to tightly bound bosonic pairs. As originally discussed in the context of continuum systems~\cite{Eagles1969,Leggett1980,Nozieres1985}, this crossover has been extensively explored in strongly interacting Fermi gases in both two and three dimensions (2D and 3D)~\cite{Greiner2003,Zwierlein2006,Schunck2008,Boronat2005,Evgeni2008,Adam2020,Feld2011,Ries2015,Murthy2015,Boettcher2016,Makhalov2014,Anderson2015,Yuanyao2022,Daix2026}, where the interaction strength can be tuned over a wide range via Feshbach resonances~\cite{Inouye1998,OHara2002,Chin2010}. Beyond continuum systems, lattice models provide a complementary setting in which the fermion pairing and associated superconductivity are controlled not only by the interaction strength but also by the lattice geometry, band structure, and fermion density. Understanding how these microscopic parameters shape the superconducting state is therefore of fundamental interest in the study of strongly correlated fermionic systems.

As a representative system, the Hubbard model~\cite{Hubbard1963,Kanamori1963,Gutzwiller1963,Arovas2022,Qin2022} with an attractive on-site interaction provides one of the simplest settings for spinful fermions to study superconductivity across the BCS-BEC crossover. This model has been investigated extensively, particularly using {\it numerically unbiased} auxiliary-field quantum Monte Carlo (AFQMC) methods~\cite{Hirsch1985,Scalettar1989,Moreo1991,Bormann1991,Scalapino1992,Scalapino1993,Assaad1993,Singer1996,Assaad1994,Paiva2004,Rodrigo2022,Wang2023,Rodrigo2024,Yuanyao2026,Nakano2006,Lee2009,Zhang2026,Rubem2026,Sewer2002,Paiva2010}, as well as by other analytical and numerical approaches~\cite{Chien2008,Tamaki2008,Kyung2001,Sakai2015,Peters2015,Karakuzu2018,Shi2023,Equbal2026,Xiong2025}. These studies have focused mostly on finite-temperature properties, including superconducting transitions~\cite{Scalettar1989,Moreo1991,Bormann1991,Assaad1994,Paiva2004,Rodrigo2022,Wang2023,Rodrigo2024,Yuanyao2026,Nakano2006,Sewer2002,Xiong2025}, thermodynamic properties~\cite{Singer1996,Lee2009,Paiva2010,Chien2008,Tamaki2008}, and pseudogap phenomena~\cite{Rodrigo2022,Wang2023,Rodrigo2024,Kyung2001,Sakai2015,Peters2015,Xiong2025}. Several studies have also established the superconducting ground state with spin-singlet $s$-wave pairing~\cite{Moreo1991,Bormann1991,Scalapino1992,Scalapino1993,Assaad1993,Karakuzu2018,Shi2023,Equbal2026}. Nevertheless, although the qualitative structure of the ground state is relatively well established, a comprehensive quantitative characterization still remains limited~\cite{Bormann1991,Scalapino1992,Scalapino1993,Assaad1993,Karakuzu2018}. In particular, systematic and high-precision benchmark results over a broad range of the interaction strength and fermion density are largely lacking, especially in 2D, where quantum fluctuations can play an important role. In this regard, the pairing order~\cite{Karakuzu2018} and condensate fraction~\cite{Boronat2005,Yuanyao2026,Yuanyao2022,Yang1962,Shihao2015,Fukushima2007,Adam2020} provide complementary measures of superconductivity, where the former characterizes the macroscopic pairing amplitude and the latter quantifies the fraction of fermion pairs occupying the Bose-condensed state. Reliable finite-size and thermodynamic-limit results for these quantities across a broad parameter regime in the attractive Hubbard model can therefore provide useful reference data for assessing approximate theoretical and numerical methods~\cite{Karakuzu2018,Shi2023,Yuntian2026}. Other fundamental physical observables, such as the double occupancy and total energy, further characterize the evolution of the superconducting ground state and can offer additional benchmarks~\cite{Shi2023}. 

In addition, next-nearest-neighbor hopping serves as another dimension for modulating the superconductivity in the attractive Hubbard model~\cite{Santos1992,Fontenele2023,Fontenele2026}. By modifying the single-particle energy dispersion and Fermi-surface structure, it can alter the kinetic properties of the system and consequently influence the formation and condensation of fermion pairs. This allows for a controlled examination of how nonlocal kinetic processes affect fermion pairing and superconductivity. Hence, a quantitative investigation of these effects can complement the ground-state characterization of the standard attractive Hubbard model, and provide further insight into the interplay between fermion pairing and single-particle band structure.

In this work, we address the above issues by systematically studying the ground state of the 2D attractive Hubbard model using the {\it numerically exact} zero-temperature AFQMC method~\cite{Sugiyama1986,Sorella1989,White1989,Assaad2008,Duhao2025}. We explore a broad and physically relevant parameter regime of the model, covering the interaction strength $U/t=2\sim 12$ and fermion density $n=0.25\sim 1.0$, thereby allowing us to characterize the superconductivity across the BCS-BEC crossover from both interaction- and density-driven perspectives. Our primary focus is on the pairing order and condensate fraction, for which we analyze their dependences on both $U/t$ and $n$, and accordingly perform finite-size extrapolations to obtain their thermodynamic-limit values. We also present results for the double occupancy and total energy as complementary benchmarks. Additionally, we examine the effects of next-nearest-neighbor hopping on fermion pairing and the superconducting ground state of the attractive Hubbard model.

The remainder of this paper is organized as follows. In Sec.~\ref{sec:modelmethod}, we describe the 2D attractive Hubbard model, the AFQMC method, and the physical observables computed in this work. In Sec.~\ref{sec:Results}, we present our AFQMC results for the standard attractive Hubbard model and its extension with next-nearest-neighbor hopping. Finally, in Sec.~\ref{sec:Summary}, we summarize our work and discuss  implications for future research. In Appendixes, we provide supplementary results on the convergence tests in AFQMC simulations and finite-size extrapolations.

\section{Model, Method and physical observables}
\label{sec:modelmethod}

\subsection{The 2D attractive Hubbard model}
\label{sec:TheModel}

We study the 2D attractive Hubbard model on a square lattice described by the following Hamiltonian
\begin{equation}\begin{aligned}
\label{eq:HubbModel}
\hat{H}
= & -t \sum_{\langle\mathbf{ij}\rangle\sigma}(c_{\mathbf{i}\sigma}^+ c_{\mathbf{j}\sigma}^{} + h.c.) 
-t_2 \sum_{\langle\langle\mathbf{ij}\rangle\rangle\sigma}(c_{\mathbf{i}\sigma}^+ c_{\mathbf{j}\sigma}^{} + h.c.) \\
&- U\sum_{\mathbf{i}}\hat{n}_{\mathbf{i}\uparrow}\hat{n}_{\mathbf{i}\downarrow},
\end{aligned}\end{equation}
where $\hat{n}_{\mathbf{i}\sigma}=c_{\mathbf{i}\sigma}^+c_{\mathbf{i}\sigma}^{}$ and $\hat{n}_{\mathbf{i}}=\hat{n}_{\mathbf{i}\uparrow}+\hat{n}_{\mathbf{i}\downarrow}$ are the density operators, with $\mathbf{i}=(i_x,i_y)$ as the coordinates of the lattice site and $\sigma$ ($=\uparrow$ or $\downarrow$) representing spin. The nearest-neighbor (NN) and next-nearest-neighbor (NNN) hopping amplitudes are denoted by $t$ and $t_2$, respectively. Under periodic boundary conditions (PBC), the corresponding kinetic energy dispersion is given by
\begin{equation}\begin{aligned}
\label{eq:EkDisp}
\varepsilon_{\mathbf{k}}
= -2t(\cos k_x + \cos k_y) - 4t_2\cos k_x \cos k_y,
\end{aligned}\end{equation}
where the lattice constant is set to unity. The momentum $k_x,k_y$ are defined in units of $2\pi/L$, and $L$ is the linear size of the square supercell with system size $N_s=L^2$ that is applied in our simulations. The fermion density reads $n=N_e/N_s$, with $N_e=N_{\uparrow}+N_{\downarrow}$ with $N_{\sigma}$ being the number of fermions in spin-$\sigma$ channel. We only consider the spin-balanced case with $N_{\uparrow}=N_{\downarrow}$. Throughout this study, we set $t$ as the energy unit and concentrate on the attractive interaction [with $U>0$ in Eq.~(\ref{eq:HubbModel})].

At half filling ($n=1$), the standard attractive Hubbard model (with $t_2=0$) possesses a charge SU(2) symmetry~\cite{Yang1989,Mark2020,Hille2020}, which guarantees the degeneracy between the charge density wave (CDW) order and on-site spin-singlet pairing order (see more details in Sec.~\ref{sec:MainResults}). Upon doping or introducing a nonzero $t_2$, this symmetry is reduced to the U(1) symmetry associated with charge conservation, by lifting the degeneracy while leaving the pairing-ordered state as the ground state~\cite{Mark2020}. Consequently, the model in Eq.~(\ref{eq:HubbModel}) exhibits a superconducting ground state with spin-singlet $s$-wave pairing long-range order for any interaction strength and fermion density~\cite{Scalettar1989,Moreo1991,Bormann1991,Singer1996}. For weak to intermediate attractive interactions, the superconductivity consists of Cooper pairs with different spatial extents, whose pairing amplitudes exhibit a characteristic oscillatory behavior as a function of the distance between the two fermions~\cite{Shihao2015,Yuanyao2026}, as described by the BCS theory~\cite{Bardeen1957a,Bardeen1957b}. Specifically, nonlocal fermion pairing in the 2D model~(\ref{eq:HubbModel}) has been observed in an optical lattice experiment~\cite{Hartke2023}. Nevertheless, owing to the local nature of the Hubbard attraction, on-site pairing order remains the dominant component of the superconducting state and has been the central focus in the attractive Hubbard model for decades~\cite{Scalettar1989,Moreo1991,Bormann1991,Singer1996,Paiva2004,Rodrigo2022,Wang2023,Rodrigo2024,Yuanyao2026,Nakano2006,Lee2009,Zhang2026,Rubem2026,Sewer2002,Chien2008,Tamaki2008,Karakuzu2018,Shi2023,Equbal2026}. Meanwhile, off-site pairing components can be manifested in the real-space pair wave function~\cite{Shihao2015,Yuanyao2026}.

\subsection{The zero-temperature AFQMC method}
\label{sec:AFQMC}

To solve the model~(\ref{eq:HubbModel}), we apply the {\it numerically exact} zero-temperature AFQMC method, also known as projector quantum Monte Carlo~\cite{Sugiyama1986,Sorella1989,White1989,Assaad2008,Duhao2025}. Here, the AFQMC simulations are free of the minus sign problem for arbitrary fermion density~\cite{Congjun2005}, and thus allow for large-scale calculations and reliable finite-size extrapolations to obtain physical observables in the thermodynamic limit (TDL). In this subsection, we briefly introduce the key ingredients of the method that are relevant to the present study.

In zero-temperature AFQMC, the ground state of a many-body system is obtained via the imaginary-time projection from a trial wave function $|\psi_T\rangle$, i.e., $|\Psi_G\rangle=\lim_{\Theta\to\infty}e^{-\frac{\Theta}{2}\hat{H}}|\psi_T\rangle$, with $\Theta$ as the projection parameter. Then the ground-state observable is calculated using
\begin{equation}\begin{aligned}
\label{eq:GSObs}
\langle\hat{O}\rangle
=\frac{\langle\Psi_{G}|\hat{O}|\Psi_{G}\rangle}{\langle\Psi_{G}|\Psi_{G}\rangle}
=\lim_{\Theta\to\infty}\frac{\langle\psi_{T}|e^{-\frac{\Theta}{2}\hat{H}}\hat{O}e^{-\frac{\Theta}{2}\hat{H}}|\psi_{T}\rangle}{\langle\psi_{T}|e^{-\Theta\hat{H}}|\psi_{T}\rangle}.
\end{aligned}\end{equation}
Provided that $|\psi_{T}\rangle$ has a nonzero overlap with the true many-body ground state, i.e., $\langle\Psi_{G}|\psi_{T}\rangle\ne0$, the projection converges to the ground state with a finite $\Theta$. Based on Eq.~(\ref{eq:GSObs}), the AFQMC method discretizes the projection as $\Theta=M\Delta\tau$ and applies the Trotter-Suzuki decomposition to $e^{-\Delta\tau\hat{H}}$. By decomposing the Hamiltonian as $\hat{H}=\hat{H}_0+\hat{H}_U$, where $\hat{H}_0$ and $\hat{H}_U$ represent the kinetic and interaction parts, respectively, we further apply the Hubbard-Stratonovich (HS) transformation to $e^{-\Delta\tau\hat{H}_U}$, thereby decoupling the two-body interactions into free fermions coupled to auxiliary fields. These procedures lead to the weight of each auxiliary-field configuration (or path) being expressed as a matrix determinant, while fermionic observables are evaluated through importance sampling over the configurations.

We then outline the important implementation details used in our AFQMC simulations. {\it First}, we choose $|\psi_{T}\rangle$ as the ground-state Slater determinant wave function of $\hat{H}_0$. The desired fermion density is imposed by setting the number of occupied single-particle orbitals to $N_{\sigma}$ (the number of spin-$\sigma$ fermions). An ambiguity in choosing the occupied orbitals arises if a degeneracy occurs in the kinetic-energy levels at the Fermi level. In this case, we include a tiny twisted boundary phase into $\hat{H}_0$ only when constructing $|\psi_{T}\rangle$~\cite{Assaad2008}, so as to lift the degeneracy and maintain the closed-shell structure. {\it Second}, we use the third-order Trotter-Suzuki decomposition for $e^{-\Delta\tau\hat{H}}$ as
\begin{equation}\begin{aligned}
\label{eq:SymTrot}
e^{-\Delta\tau\hat{H}}=e^{-\Delta\tau\hat{H}_0/2}e^{-\Delta\tau\hat{H}_U}e^{-\Delta\tau\hat{H}_0/2}+\mathcal{O}[(\Delta\tau)^3],
\end{aligned}\end{equation}
which generally yields an overall Trotter error $\propto(\Delta\tau t)^2$ for physical observables in AFQMC calculations. It has been shown~\cite{Song2025B} that this symmetric formula leads to substantially faster convergence toward the $\Delta\tau t\to0$ limit than the asymmetric one. We further eliminate the residual Trotter error by performing convergence tests with varying $\Delta\tau t$ until the results become statistically indistinguishable within the estimated uncertainties. We have also carried out similar convergence tests for the projection parameter $\Theta t$. The results of both convergence tests, together with the simulation parameters $\Delta\tau t$ and $\Theta t$ adopted in this work, are summarized in Appendix~\ref{sec:ConvergeTest}. {\it Third}, we apply the HS transformation into the spin-$\hat{s}^z$ channel with two-component auxiliary fields for the Hubbard interaction~\cite{Hirsch1983} as
\begin{equation}\begin{aligned}
\label{eq:HSspinDecomp}
e^{+\Delta\tau U \big(\hat{n}_{\mathbf{i}\uparrow} \hat{n}_{\mathbf{i}\downarrow} - \frac{\hat{n}_{\mathbf{i}\uparrow} + \hat{n}_{\mathbf{i}\downarrow}}{2}\big) }
= \frac{1}{2}\sum_{x_{\mathbf{i}}=\pm1}e^{i\gamma x_{\mathbf{i}}(\hat{n}_{\mathbf{i}\uparrow}-\hat{n}_{\mathbf{i}\downarrow})},
\end{aligned}\end{equation}
where $\gamma=\cos^{-1}(e^{-\Delta\tau U/2})$ is the coupling coefficient for $U>0$ in the model~(\ref{eq:HubbModel}), and $x_{\mathbf{i}}=\pm1$ is the auxiliary field. This transformation has been shown to produce high-precision results for density- and pairing-related quantities~\cite{Yuanyao2026,Song2025B,Zhuotao2025}, though it results in the complex-number framework for AFQMC simulations. {\it Fourth}, we implement the fast Fourier transform (FFT)~\cite{Yuanyao2026,Houmin2026} in the path propagation and the delayed update algorithm~\cite{Duhao2025} for the configuration update process. The former accelerates the matrix-matrix multiplication in the propagation, and the latter is an improved version of the conventional local update. The combined use of these two techniques can yield substantial speedups in AFQMC simulations~\cite{Duhao2025,Houmin2026}, with the speedup growing with system size. Moreover, the FFT approach facilitates easy access to the momentum-space Green's function, and thus allows us to directly measure related quantities in AFQMC, such as the momentum-space pairing matrix and associated condensate fraction (see Sec.~\ref{sec:Observables}). {\it Fifth}, we perform multiple measurements of physical observables for a selected configuration within an imaginary-time interval, i.e., $0.45\Theta\le\tau\le0.55\Theta$, rather than performing a single measurement strictly at the midpoint as specified in Eq.~(\ref{eq:GSObs}). This is reasonable and valid as long as the projection time remaining ($0.45\Theta$ in our simulations) on both sides of the measurement interval is sufficiently long to ensure convergence to the ground state. {\it Sixth}, we mostly impose PBC in the simulations, and also apply twist-averaged boundary conditions (TABC)~\cite{Song2025B,Houmin2026} for weak interactions where finite-size effects are significant. Together, these ingredients provide a highly efficient AFQMC framework to simulate the model~(\ref{eq:HubbModel}) and enable us to reach high-precision numerical results. Furthermore, we note that the zero-temperature AFQMC method can suffer from the infinite variance problem~\cite{Shihao2016Inf,Maksim2022,Yunus2022,Andrei2023,Wan2025}, even in sign-problem-free simulations. Nevertheless, a generally applicable approach to completely eliminate this issue for all physical observables is still lacking~\cite{InfVarIssue}. Therefore, we do not specifically address this issue in our AFQMC simulations in the present work. For further implementation details of the AFQMC method, we refer the reader to our previous work in Ref.~\cite{Duhao2025} and the review in Ref.~\cite{Assaad2008}.

\subsection{Physical observables}
\label{sec:Observables}

In this subsection, we describe the observables measured in our AFQMC calculations for the model~(\ref{eq:HubbModel}). We focus on pairing-related observables, including the pair-pair correlation function and condensate fraction, as well as commonly used quantities such as the double occupancy, total energy, and fermion momentum distribution.

For the fermion pairing, we consider the on-site spin-singlet pairing order $\hat{\Delta}_{\mathbf{i}}^+=c_{\mathbf{i}\uparrow}^+c_{\mathbf{i}\downarrow}^+$, and accordingly measure its real-space correlation function using
\begin{equation}\begin{aligned}
\label{eq:PairCrFt}
P(\mathbf{r})=
\frac{1}{4N_s}\sum_{\mathbf{i}}\big\langle\hat{\Delta}_{\mathbf{i}}^{+}\hat{\Delta}_{\mathbf{i}+\mathbf{r}}^{} + \hat{\Delta}_{\mathbf{i}}^{}\hat{\Delta}_{\mathbf{i}+\mathbf{r}}^{+}\big\rangle,
\end{aligned}\end{equation}
for which the corresponding pairing structure factor is expressed as $S_{\rm pair}(\mathbf{q})=\sum_{\mathbf{r}}P(\mathbf{r})e^{i\mathbf{q}\cdot\mathbf{r}}$. From $S_{\rm pair}(\mathbf{q})$, we can compute the mean-squared pairing order $\langle\Delta^2\rangle$ using $\langle\Delta^2\rangle=S_{\rm pair}(\mathbf{q}=\boldsymbol{\Gamma})/N_s$. At TDL, the superconducting order parameter is given by $\Delta_s=\lim_{L\to\infty}\sqrt{\langle\Delta^2\rangle}$. An alternative way to obtain $\Delta_s$ is via the correlation $P(\mathbf{r})$ evaluated at large and fixed $\alpha=|\mathbf{r}|/L$, using the relation $\Delta_s=\lim_{L\to\infty}\sqrt{P(|\mathbf{r}|=\alpha L)}$. In this work, we choose $\mathbf{r}=(L/2,0)$ and $\mathbf{r}=(L/2,L/2)$ for such calculations. Consequently, as $L\to\infty$, $P(|\mathbf{r}|=\alpha L)$ and $\langle\Delta^2\rangle$ should extrapolate to the same TDL value, which provides a useful consistency check on the reliability of our finite-size extrapolations. Besides, we also calculate the vertex contribution for $P(\mathbf{r})$, denoted as $\overline{P}(\mathbf{r})$, by subtracting the uncorrelated pair correlation~\cite{White1989b,Zhang1997}.

We also probe the pairing from the momentum-space perspective via the pairing matrix~\cite{Shihao2015,YuanYao2019,Yuanyao2022,Yuanyao2026}
\begin{equation}\begin{aligned}
\label{eq:PairMat}
M_{\mathbf{k}\mathbf{k}^{\prime}}
=\langle\hat{\Delta}_{\mathbf{k}}^+\hat{\Delta}_{\mathbf{k}^{\prime}}^{}\rangle-\delta_{\mathbf{k}\mathbf{k}^{\prime}}\langle c_{\mathbf{k}\uparrow}^+c_{\mathbf{k}\uparrow}^{}\rangle\langle c_{-\mathbf{k}\downarrow}^+c_{-\mathbf{k \downarrow}^{}}\rangle,
\end{aligned}\end{equation}
where $\hat{\Delta}_{\mathbf{k}}^+=c_{\mathbf{k}\uparrow}^+c_{-\mathbf{k}\downarrow}^+$ is the $\mathbf{k}$-space spin-singlet pairing operator with {\it zero center-of-mass momentum}. The above pairing matrix is the dominant component of the real-space two-body density matrix $\langle c_{\mathbf{i}\uparrow}^+c_{\mathbf{j}\downarrow}^+c_{\mathbf{i}+\mathbf{r}\uparrow}^{}c_{\mathbf{j}+\mathbf{r}^{\prime}\downarrow}^{}\rangle$ related to the off-diagonal long-range order~\cite{Yang1962,Boronat2005} for the attractive Hubbard model. From the leading eigenvalue of $M_{\mathbf{k}\mathbf{k}^{\prime}}$ (denoted $\lambda_{\rm max}$), we obtain the condensate fraction as $n_c=\lambda_{\rm max}/(N_e/2)$, meaning the fraction of Bose-condensed fermion pairs. The corresponding eigenstate of $\lambda_{\rm max}$ is the $\mathbf{k}$-space pair wave function $\phi_{\uparrow\downarrow}(\mathbf{k})$, whose squared modulus identifies the probability of fermions with momentum $\mathbf{k}$ participating in pairing. The $\mathbf{r}$-space pair wave function $\psi_{\uparrow\downarrow}(\mathbf{r})$ can be computed via a Fourier transform, i.e., $\psi_{\uparrow\downarrow}(\mathbf{r})=(1/\sqrt{N_s})\sum_{\mathbf{k}}e^{-i\mathbf{k}\cdot\mathbf{r}}\phi_{\uparrow\downarrow}(\mathbf{k})$, and $|\psi_{\uparrow\downarrow}(\mathbf{r})|^2$ quantifies the probability of fermion pairing with a distance $\mathbf{r}$ between the two fermions. Both $\phi_{\uparrow\downarrow}(\mathbf{k})$ and $\psi_{\uparrow\downarrow}(\mathbf{r})$ can serve as important indicators of the BCS-BEC crossover~\cite{Shihao2015,Yuanyao2022}. In our AFQMC calculations, we directly measure $M_{\mathbf{k}\mathbf{k}^{\prime}}$ for a specific configuration by applying the Wick decomposition from the $\mathbf{k}$-space single-particle Green's function as $\mathbf{G}^{\sigma}=\{G_{\mathbf{k},\mathbf{k}^{\prime}}^{\sigma}=\langle c_{\mathbf{k}\sigma}^{}c_{\mathbf{k}^{\prime}\sigma}^+\rangle_{\tau}\}$, which is calculated via FFT of the $\mathbf{r}$-space Green's function matrix. 

For the commonly used observables, we compute the double occupancy $D=N_s^{-1}\sum_{\mathbf{i}}\hat{n}_{\mathbf{i}\uparrow}\hat{n}_{\mathbf{i}\downarrow}$ and the total energy density $e=\langle\hat{H}\rangle/N_s$. We also calculate the fermion momentum distribution using $n(\mathbf{k})=\sum_{\mathbf{r}}G(\mathbf{r})e^{i\mathbf{k}\cdot\mathbf{r}}$ with $G(\mathbf{r})=(2N_s)^{-1}\sum_{\mathbf{i}\sigma}\langle c_{\mathbf{i},\sigma}^{+}c_{\mathbf{i}+\mathbf{r},\sigma}^{}\rangle$ being the $\mathbf{r}$-space single-particle Green's function. The evolution of $n(\mathbf{k})$ can facilitate the visualization of the BCS-BEC crossover~\cite{Shihao2015}.

In this work, we obtain AFQMC results of the above physical quantities for the standard Hubbard model~(\ref{eq:HubbModel}) (with $t_2=0$) on finite-size systems with $L=4\sim 24$, and further extrapolate the results to the TDL. Our calculations cover a broad parameter regime with $U/t=2\sim12$ and $n=0.25\sim1.0$. For the case with NNN hopping, we focus on $U/t=6$, $n=0.875$, and $-0.45\le t_2/t\le0.45$ to examine the effects of a nonzero $t_2$. For the two sets of calculations, we summarize both the finite-size and TDL data for $\langle\Delta^2\rangle$, $n_c$, $D$, and $e/t$ in Ref.~\cite{HubbData2026}.

\section{Numerical results}
\label{sec:Results}

In this section, we present our AFQMC simulation results for the model~(\ref{eq:HubbModel}). In Secs.~\ref{sec:QmcEdCompare}--\ref{sec:MainResults}, we concentrate on the standard attractive Hubbard model (with $t_2=0$), covering a benchmark comparison with Exact Diagonalization (ED), a characterization of the BCS-BEC crossover, and our main numerical results. In Sec.~\ref{sec:t2Hubbard}, we show the results for the model with nonzero $t_2$ and discuss the underlying physics. 

\begin{figure}[t]
\centering
\includegraphics[width=0.987\columnwidth]{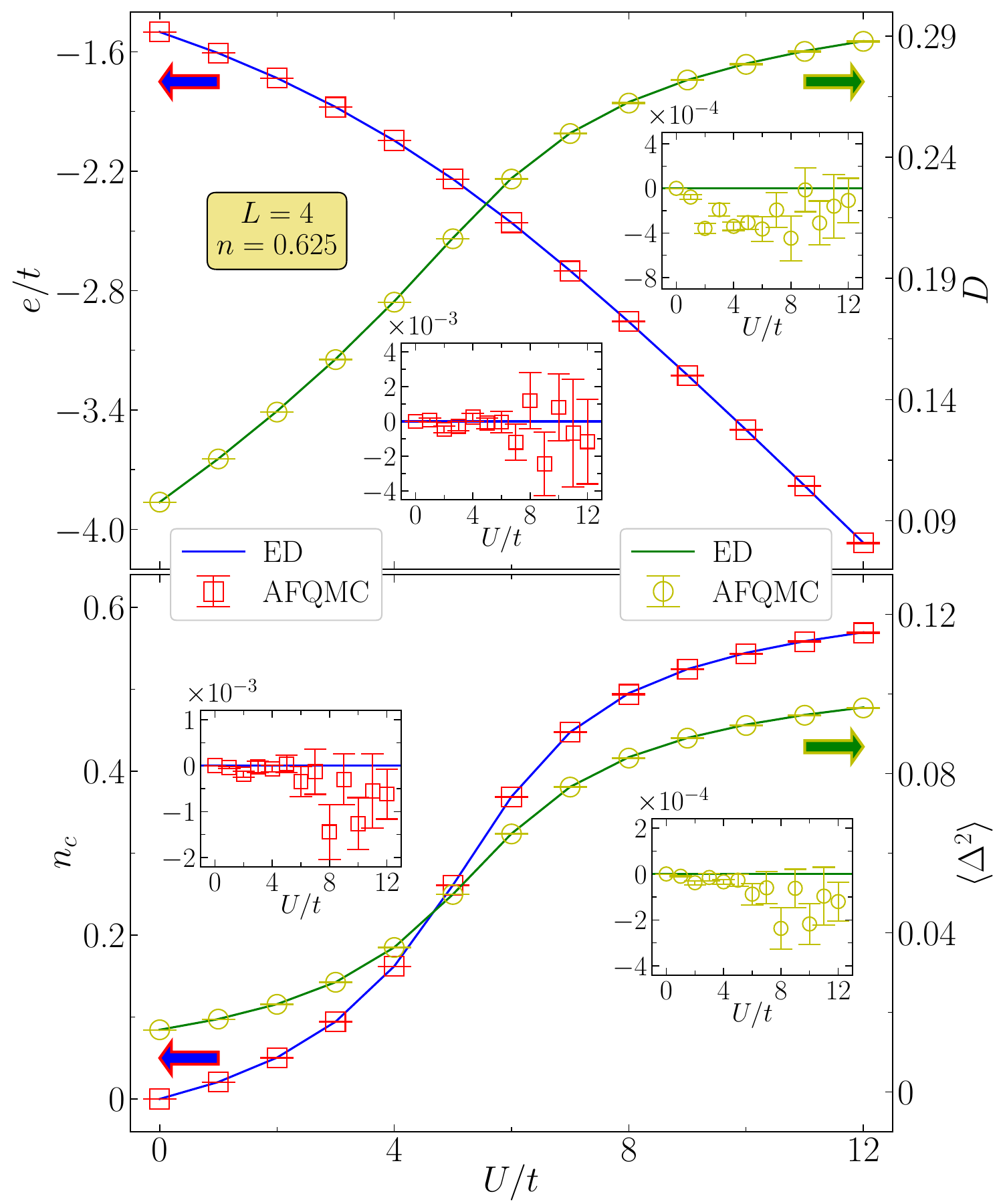}
\caption{
Benchmark comparison between AFQMC and ED for (a) the total energy density $e/t$ and double occupancy $D$, and (b) the mean-squared pairing order $\langle \hat{\Delta}^2 \rangle$ and condensate fraction $n_c$, as functions of the interaction strength $U/t$ for the standard 2D attractive Hubbard model~(\ref{eq:HubbModel}) on an $L=4$ system with PBC. The fermion density is $n=0.625$. The insets show the differences between the AFQMC and ED results for the corresponding observables, with ED taken as the reference.
}
\label{fig:fig01QmcED}
\end{figure}

\begin{figure*}[t]
\centering
\includegraphics[width=1.00\linewidth]{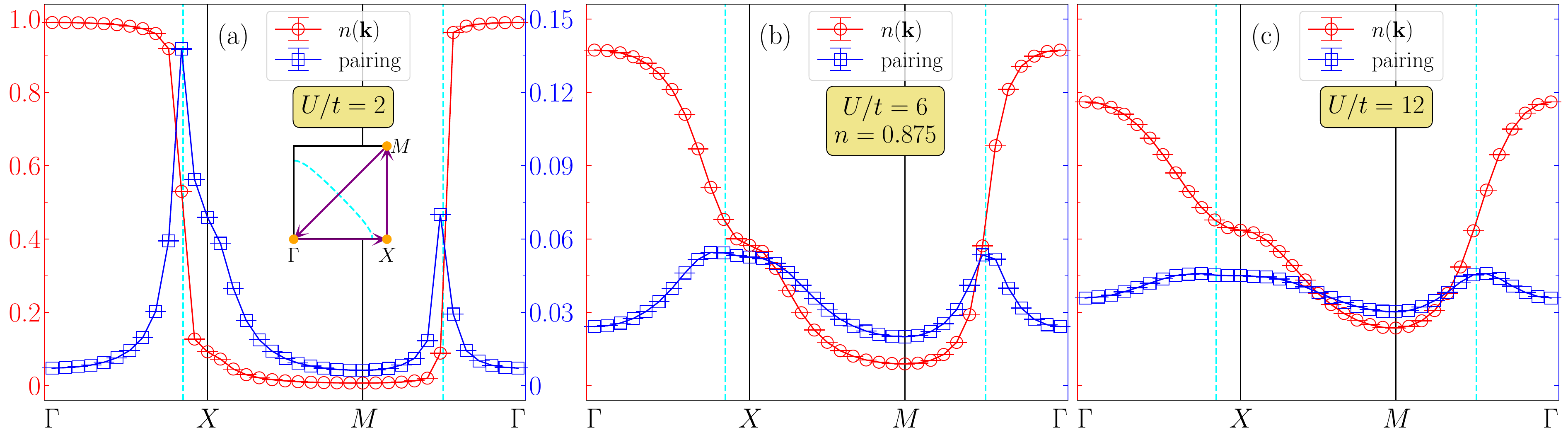}
\caption{
Momentum distribution $n(\mathbf{k})$ and $\mathbf{k}$-space pair wave function $\phi_{\uparrow\downarrow}(\mathbf{k})$ along the high-symmetry path $\Gamma\to X\to M\to\Gamma$ in the Brillouin zone for three interaction strengths: (a) $U/t=2$, (b) $U/t=6$, and (c) $U/t=12$. The fermion density is $n=0.875$ in all three panels. The left and right $y$-axis scales and tick labels are identical for all panels. Vertical cyan dashed lines indicate the noninteracting Fermi surface, which is shown in the inset of panel (a) within the quarter Brillouin zone. The calculations are performed with $L=24$ under PBC for the standard 2D attractive Hubbard model~(\ref{eq:HubbModel}).
}
\label{fig:fig02PairWFnK}
\end{figure*}

\subsection{Benchmark against Exact Diagonalization}
\label{sec:QmcEdCompare}

We first perform a sanity check for our ground-state AFQMC simulations for the standard 2D attractive Hubbard model~(\ref{eq:HubbModel}) via the benchmark with ED on an $L=4$ system. In Fig.~\ref{fig:fig01QmcED}, we plot the comparison results for the total energy density $e/t$, double occupancy $D$, condensate fraction $n_c$, and mean-squared pairing order $\langle\Delta^2\rangle$ as functions of interaction strength $U/t$ at fixed fermion density $n=0.625$. Excellent agreement is observed between the two methods for all four quantities, as evidenced by the AFQMC data points lying essentially on top of the ED curves. Their differences shown in the insets are mostly consistent with zero within the statistical uncertainties of the AFQMC data, despite the latter growing with $U/t$ as expected. Among these quantities, $D$ typically features particularly small error bars (with relative errors $<10^{-4}$) in AFQMC using the HS-$\hat{s}^z$ transformation [Eq.~(\ref{eq:HSspinDecomp})] at weak to intermediate interactions, which makes the Trotter error clearly visible, as illustrated by the deviations from the ED results at $U/t=2\sim 6$ in the right-side inset of Fig.~\ref{fig:fig01QmcED}(a). This issue was also observed in Ref.~\cite{Yufeng2026} and can be eliminated by decreasing $\Delta\tau t$. For $U/t>6$, the improved consistency with ED is attributed to the smaller $\Delta\tau t$ used at larger $U/t$ (see Appendix~\ref{sec:ConvergeTest}) and the increasingly dominant statistical errors. These benchmark results against ED validate the reliability of our AFQMC method and provide a solid foundation for extending the simulations of the model~(\ref{eq:HubbModel}) to larger system sizes.

\subsection{Characterization of the BCS-BEC Crossover}
\label{sec:BCSBEC}

\begin{figure*}[t]
\centering
\includegraphics[width=1.00\linewidth]{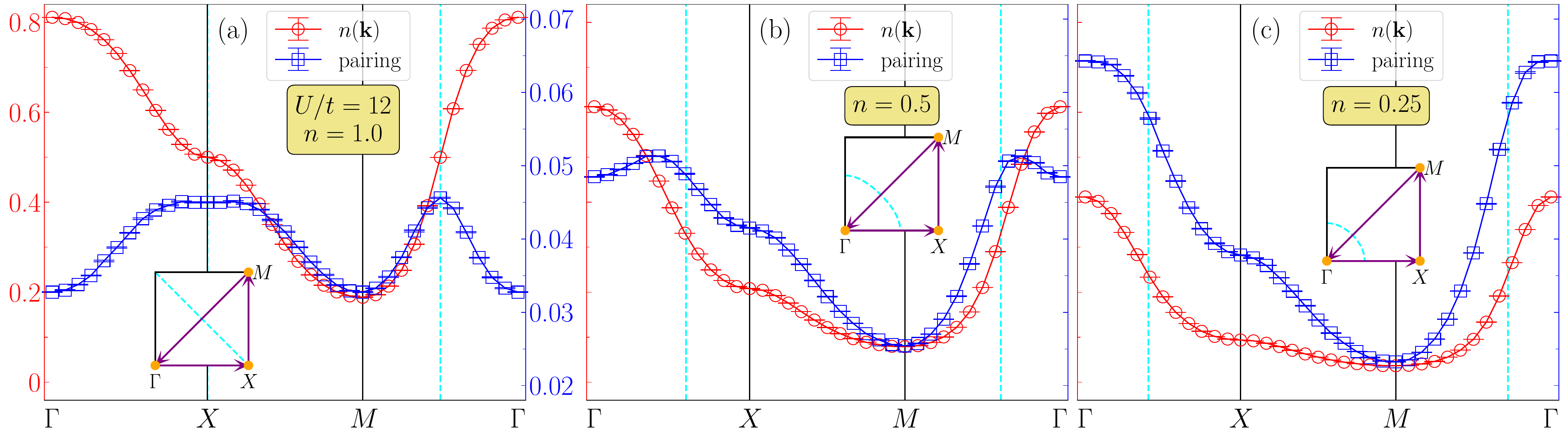}
\caption{
Momentum distribution $n(\mathbf{k})$ and $\mathbf{k}$-space pair wave function $\phi_{\uparrow\downarrow}(\mathbf{k})$ along the high-symmetry path $\Gamma\to X\to M\to\Gamma$ in the Brillouin zone for three fermion densities: (a) $n=1.0$, (b) $n=0.5$, and (c) $n=0.25$. The interaction strength is $U/t=12$ in all three panels. The left and right $y$-axis scales and tick labels are identical for all panels. Vertical cyan dashed lines indicate the noninteracting Fermi surface, which is shown in the inset of each panel within the quarter Brillouin zone. The calculations are performed with $L=24$ under PBC for the standard 2D attractive Hubbard model~(\ref{eq:HubbModel}).
}
\label{fig:fig03PairWFnK}
\end{figure*}

\begin{figure}[b]
\centering
\includegraphics[width=0.940\columnwidth]{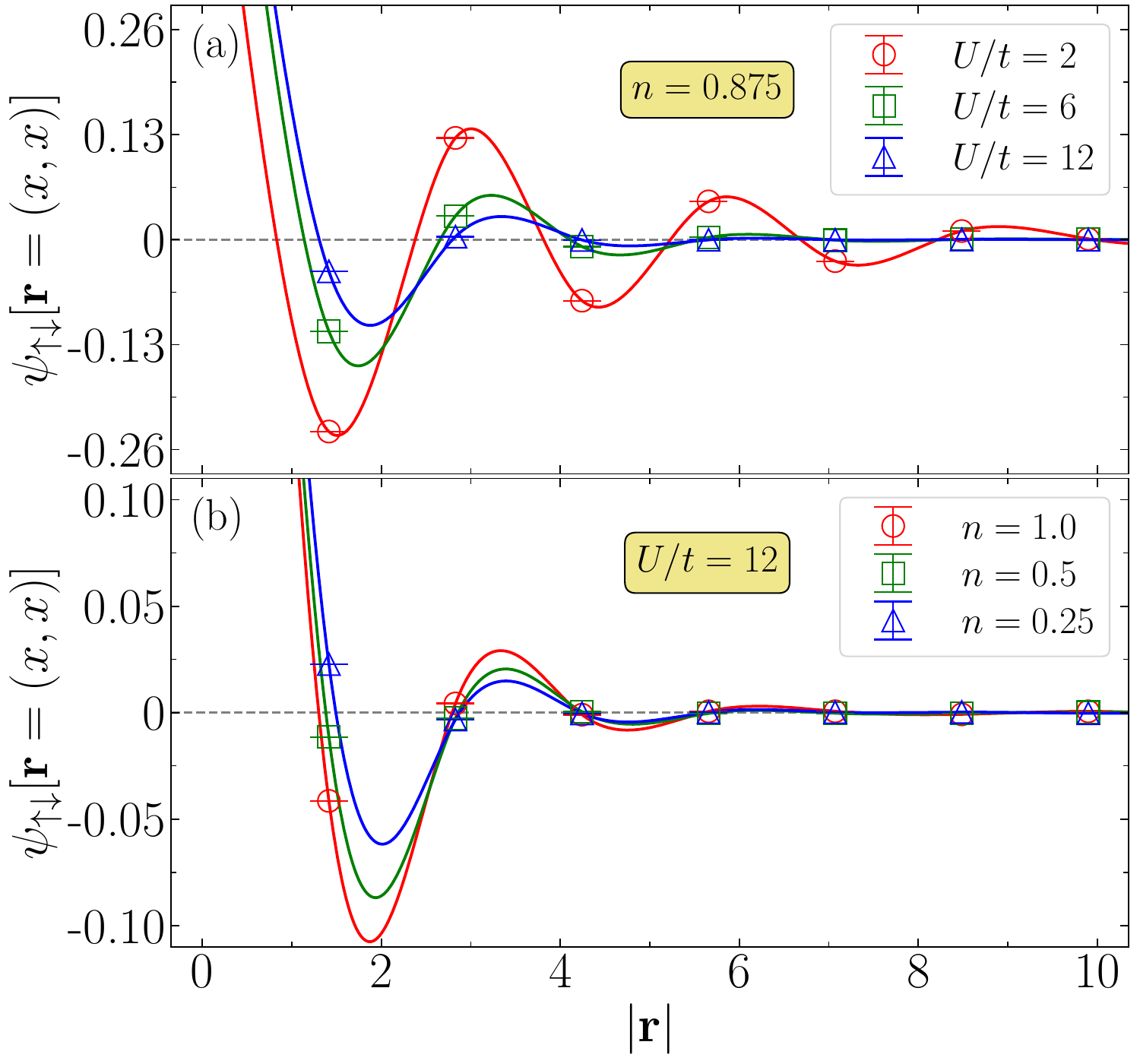}
\caption{
The $\mathbf{r}$-space pair wave function $\psi_{\uparrow\downarrow}(\mathbf{r})$ as a function of $|\mathbf{r}|$ along the diagonal direction [$\mathbf{r}=(x,x)$] for (a) $n=0.875$ with $U/t=2,6,12$, and (b) $U/t=12$ with $n=1.0,0.5,0.25$. The solid lines are spline interpolations of the AFQMC data points. Panels (a) and (b) correspond to the results shown in Figs.~\ref{fig:fig02PairWFnK} and~\ref{fig:fig03PairWFnK}, respectively, and are also obtained with $L=24$.
}
\label{fig:fig04rPairWvfc}
\end{figure}

Phenomenologically, the BCS-BEC crossover is often characterized by the size of Cooper pairs~\cite{Stefano2008,Leggett2012,Mohit2014,Giancarlo2018,Chen2005,Chen2024}, which evolves continuously from strongly overlapping pairs on the BCS side to tightly bound molecules on the BEC side. However, it is rather challenging to directly access the Cooper pair size for a superconducting state. An alternative and physically intuitive picture of the crossover is provided by the competition between the pair-binding energy $\epsilon_b$ and Fermi energy $E_F$~\cite{Shihao2015}, where the crossover is accompanied by a progressive increase of $\epsilon_b/E_F$ from $\epsilon_b/E_F\ll 1$ in the BCS regime to $\epsilon_b/E_F\gg 1$ in the BEC regime. In the context of the lattice Hubbard model, $\epsilon_b$ is predominantly determined by the interaction strength $U/t$, and $E_F$ is the Fermi energy measured relative to the bottom of the noninteracting band and primarily depends on the fermion density $n$. Specifically, $E_F$ vanishes in the dilute limit, i.e., $E_F\to0$ as $n\to0$, consistent with the situation in a 2D Fermi gas, where $E_F\propto n$~\cite{Makhalov2014,Shihao2015,Yuanyao2022}.
Thus, this energetic picture suggests that both increasing $U/t$ at fixed $n$ and decreasing $n$ at fixed $U/t$ can drive the system toward the BEC regime. In this subsection, we therefore characterize the resulting BCS-BEC crossover in the attractive Hubbard model through the evolution of the $\mathbf{k}$- and $\mathbf{r}$-space pair wave functions $\phi_{\uparrow\downarrow}(\mathbf{k})$ and $\psi_{\uparrow\downarrow}(\mathbf{r})$ as well as the fermion momentum distribution $n(\mathbf{k})$, following the approach previously employed for 2D interacting Fermi gases~\cite{Shihao2015,Yuanyao2022}.

The most widely studied BCS-BEC crossover is driven by the interaction strength~\cite{Boronat2005,Evgeni2008,Adam2020,Feld2011,Ries2015,Murthy2015,Boettcher2016,Makhalov2014,Anderson2015,Yuanyao2022,Daix2026}. In the attractive Hubbard model, this can be understood energetically as promoting the ratio $\epsilon_b/E_F$ by primarily enhancing $\epsilon_b$ with increasing $U/t$. In Fig.~\ref{fig:fig02PairWFnK}, we illustrate this crossover by presenting AFQMC results of $\phi_{\uparrow\downarrow}(\mathbf{k})$ and $n(\mathbf{k})$ for the model~(\ref{eq:HubbModel}) at $U/t=2,6,12$ with $n=0.875$. At weak interaction strength  $U/t=2$, $\phi_{\uparrow\downarrow}(\mathbf{k})$ exhibits a sharp peak near the Fermi surface, indicating that fermions in the vicinity of the Fermi surface predominantly participate in pairing. This is consistent with Cooper's original picture of pair formation~\cite{Cooper1956} and can be regarded as a prominent feature of the BCS regime. As $U/t$ increases, the peak near the Fermi surface in $\phi_{\uparrow\downarrow}(\mathbf{k})$ is gradually suppressed with the weight spreading to other momenta, reflecting the increasingly extended distribution of $\mathbf{k}$-space pairing, as illustrated by the results for $U/t=6$. Then, at strong coupling $U/t=12$, $\phi_{\uparrow\downarrow}(\mathbf{k})$ is rather flat across the entire Brillouin zone, with the peak becoming strongly broadened and barely visible. This means that the fermion pairing is broadly distributed in momentum space, with fermions of different momenta participating in pairing with comparable weights. By the Fourier transform, such a broad $\mathbf{k}$-space distribution corresponds to a spatially localized pairing structure in real space, thereby signaling the BEC regime. Turning to $n(\mathbf{k})$, we observe a step-like feature near the Fermi surface at $U/t=2$, which resembles that of the noninteracting system and is also characteristic of the BCS regime. With increasing $U/t$, this feature becomes less prominent and continuously evolves into a smooth profile near the Fermi surface. Eventually, at $U/t=12$, $n(\mathbf{k})$ becomes rather smooth throughout the Brillouin zone, showing no pronounced signature of the Fermi surface. This evolution is consistent with the increasingly localized nature of the fermion pairing inferred from $\phi_{\uparrow\downarrow}(\mathbf{k})$, and provides a complementary signature of the crossover toward the BEC regime.

Another less extensively studied BCS-BEC crossover is driven by the fermion density~\cite{Andrenacci1999,Erhai2006,Yuhao2024}. For the Hubbard model, this crossover corresponds to promoting $\epsilon_b/E_F$ by primarily reducing $E_F$ with decreasing fermion density $n$, when the interaction $U/t$ is fixed. In Fig.~\ref{fig:fig03PairWFnK}, we present the AFQMC results of $\phi_{\uparrow\downarrow}(\mathbf{k})$ and $n(\mathbf{k})$ for the model~(\ref{eq:HubbModel}) at $n=1.0$, $0.5$, and $0.25$ with $U/t=12$. For these three densities, $E_F/t$ decreases from $4.0$ at $n=1.0$, to $\simeq2.560$ at $n=0.5$, and $\simeq1.421$ at $n=0.25$. We find that $\phi_{\uparrow\downarrow}(\mathbf{k})$ and $n(\mathbf{k})$ exhibit similar, though less pronounced, signatures of the crossover to those observed in Fig.~\ref{fig:fig02PairWFnK}. Here, the peak feature near the Fermi surface in $\phi_{\uparrow\downarrow}(\mathbf{k})$ is relatively visible at $n=1.0$ (particularly along the $M\to\Gamma$ segment), becomes rather weak at $n=0.50$, and eventually disappears at $n=0.25$, where $\phi_{\uparrow\downarrow}(\mathbf{k})$ varies smoothly and monotonically across the Fermi surface. Note that the broader $y$-axis range of $\phi_{\uparrow\downarrow}(\mathbf{k})$ at lower densities essentially arises from the shrinking Fermi surface and the reduced occupation of fermions with momenta near the corners of the Brillouin zone at the $U/t$ considered here. Regarding $n(\mathbf{k})$, although it exhibits smooth behavior for all three densities, as $n$ decreases, its most rapid variation across the Fermi surface is progressively less prominent and its $y$-axis range becomes narrower. Moreover, $n(\mathbf{k})$ is continuously suppressed around the Brillouin zone center (the $\Gamma$ point). These observations indicate that the effective interaction effects become stronger as $n$ decreases (though $U/t=12$ is fixed in Fig.~\ref{fig:fig03PairWFnK}), and are therefore consistent with the crossover to the BEC regime. 

Besides $\phi_{\uparrow\downarrow}(\mathbf{k})$ and $n(\mathbf{k})$, the BCS-BEC crossover also manifests in the evolution of the $\mathbf{r}$-space pair wave function $\psi_{\uparrow\downarrow}(\mathbf{r})$. In Fig.~\ref{fig:fig04rPairWvfc}, we show the corresponding $\psi_{\uparrow\downarrow}(\mathbf{r})$ results [along $\mathbf{r}=(x,x)$ direction] for the same parameters as in Figs.~\ref{fig:fig02PairWFnK} and~\ref{fig:fig03PairWFnK}, obtained from the same AFQMC simulations. These results clearly demonstrate the dominant contribution from local pairing around $|\mathbf{r}|\to0$ (not plotted explicitly due to the $y$-axis scale), as expected for the on-site interaction. For the interaction-driven BCS-BEC crossover [see Fig.~\ref{fig:fig04rPairWvfc}(a)], $\psi_{\uparrow\downarrow}(\mathbf{r})$ displays an oscillating wave structure, which extends to $|\mathbf{r}|$ comparable to the largest distance in the finite-size system at $U/t=2$. This reveals the extended $\mathbf{r}$-space fermion pairing in the BCS regime. As $U/t$ increases, the wave structure is progressively suppressed, and $\psi_{\uparrow\downarrow}(\mathbf{r})$ gradually shrinks to $|\mathbf{r}|\to0$, indicating the evolution of the pairing toward tightly bound molecules in the BEC regime. As for the density-induced crossover [see Fig.~\ref{fig:fig04rPairWvfc}(b)], $\psi_{\uparrow\downarrow}(\mathbf{r})$ shows a similar suppression of the wave structure with decreasing $n$, consistent with the evolution toward the BEC regime. 

In Figs.~\ref{fig:fig02PairWFnK}--\ref{fig:fig04rPairWvfc}, we illustrate the BCS-BEC crossovers in the standard attractive Hubbard model~(\ref{eq:HubbModel}) at selected model parameters. We verify that the interaction-driven crossover also occurs at other fermion densities, whereas the density-induced crossover appears to persist only for sufficiently strong interactions (e.g., for $U/t\ge6$, from our results of condensate fraction; see Sec.~\ref{sec:MainResults}). For the latter, as a comparison, we also present AFQMC results of $n(\mathbf{k})$, $\phi_{\uparrow\downarrow}(\mathbf{k})$, and $\psi_{\uparrow\downarrow}(\mathbf{r})$ at $n=1.0$, $0.5$, and $0.25$ for $U/t=6$ in Appendix~\ref{sec:U6densityCross}. These results exhibit weak but observable signatures of the density-induced crossover.

\subsection{Finite-size and TDL results }
\label{sec:MainResults}

\begin{figure}[t]
\centering
\includegraphics[width=1.00\columnwidth]{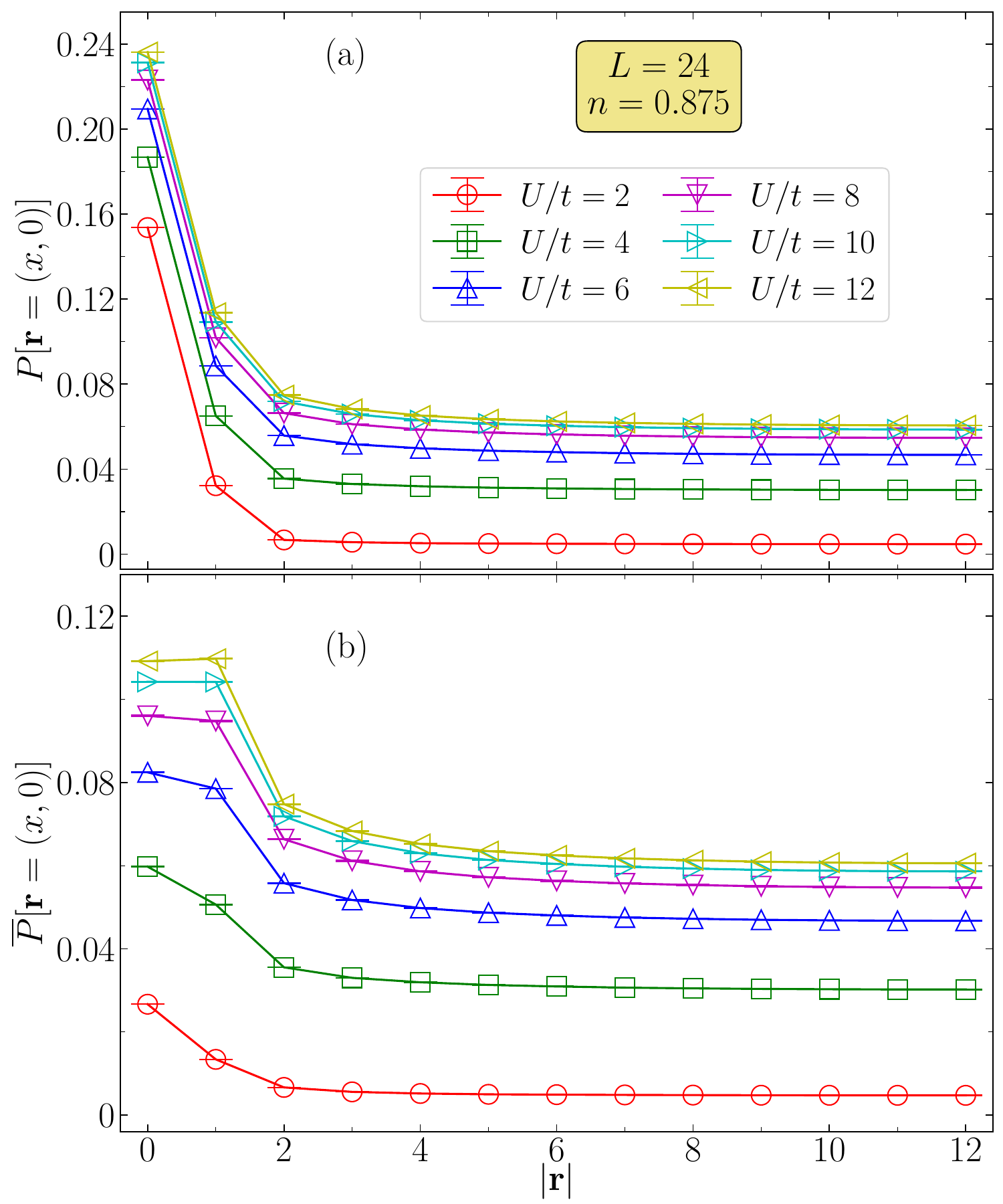}
\caption{
Pair-pair correlation functions as a function of distance $|\mathbf{r}|$ along the $x$ direction [$\mathbf{r}=(x,0)$] for the standard 2D attractive Hubbard model~(\ref{eq:HubbModel}) with $L=24$ under PBC at $n=0.875$, for interaction strength $U/t=2\sim 12$. The total pair-pair correlation function $P(\mathbf{r})$ and its vertex contribution $\overline{P}(\mathbf{r})$ are shown in (a) and (b), respectively.
}
\label{fig:fig05PairCrFt}
\end{figure}

\begin{figure}[t]
\centering
\includegraphics[width=0.988\columnwidth]{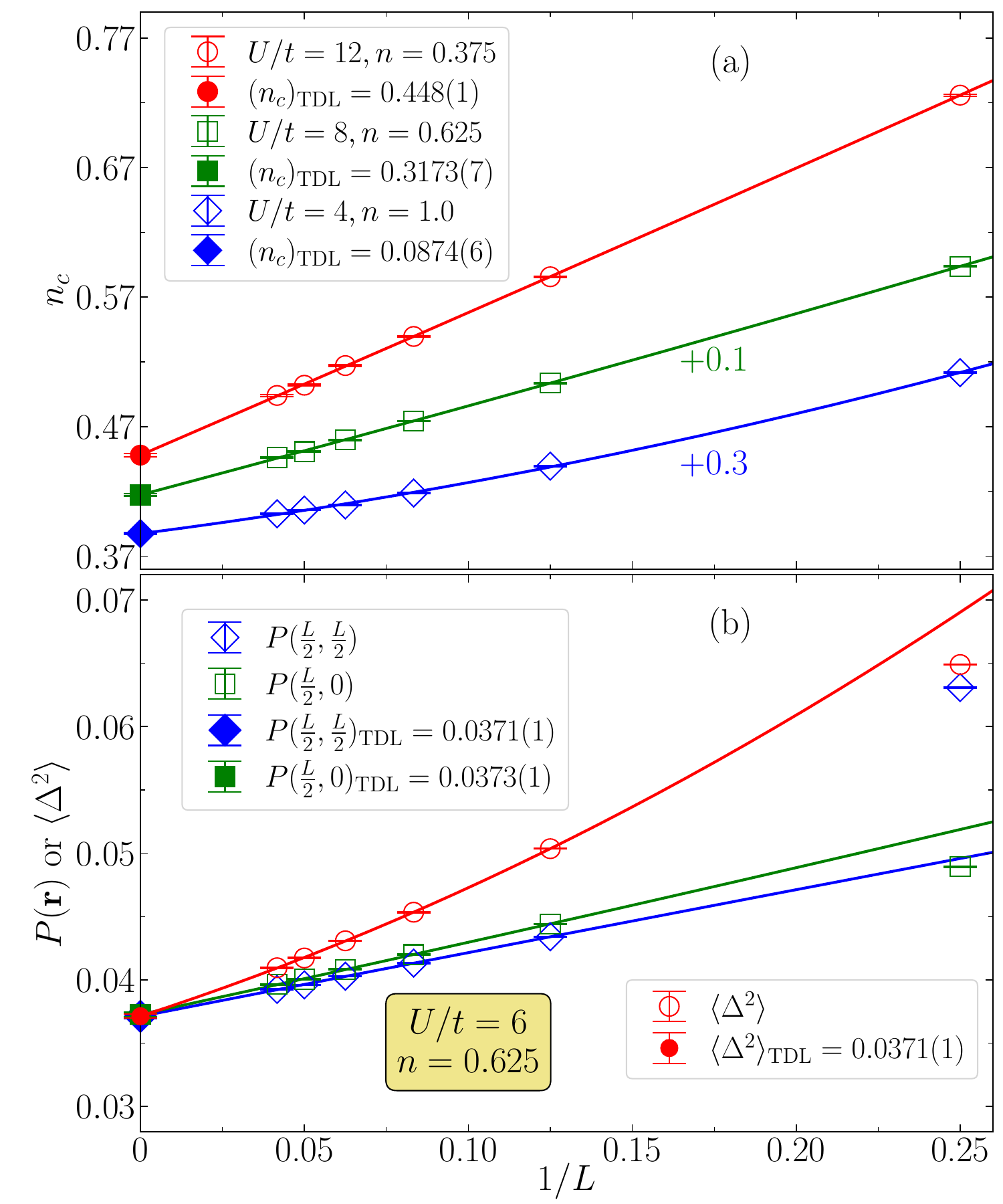}
\caption{
Finite-size extrapolations of (a) condensate fraction $n_c$, and (b) mean-squared pairing order $\langle\Delta^2\rangle$ and pair-pair correlations $P[\mathbf{r}=(L/2,L/2)]$ and $P[\mathbf{r}=(L/2,0)]$, versus $1/L$ for the standard 2D attractive Hubbard model~(\ref{eq:HubbModel}) under PBC. In panel (a), three sets of parameters are shown: $(U/t=12,n=0.375)$, $(U/t=8,n=0.625)$, and $(U/t=4,n=1.0)$, with the data for the latter two shifted by $+0.1$ and $+0.3$, respectively, for clarity. In panel (b), the parameter set $(U/t=6,n=0.625)$ is used. The solid lines represent least-squares fits to the AFQMC data using quadratic polynomials in $1/L$. The corresponding TDL values obtained from the extrapolations are also included.
}
\label{fig:fig06Extrapolate}
\end{figure} 

In this subsection, we concentrate on our main numerical results for the standard Hubbard model~(\ref{eq:HubbModel}) across the interaction-driven and density-induced BCS-BEC crossovers. These include the pair-pair correlation $P(\mathbf{r})$, finite-size extrapolations of the condensate fraction $n_c$ and mean-squared pairing order $\langle\Delta^2\rangle$, the pairing order at half filling, and TDL results for $n_c$, $\langle\Delta^2\rangle$, the total energy density $e/t$ and double occupancy $D$. Together, these results provide further insights into the BCS-BEC crossover physics in the attractive Hubbard model. 

We begin with representative results for the correlation $P(\mathbf{r})$ and its vertex contribution $\overline{P}(\mathbf{r})$. In Fig.~\ref{fig:fig05PairCrFt}, we plot these correlations as functions of $|\mathbf{r}|$ [along $\mathbf{r}=(x,0)$ direction] for $U/t=2\sim12$ at $n=0.875$ on a system with $L=24$. For all $U/t$, both $P(\mathbf{r})$ and $\overline{P}(\mathbf{r})$ first decrease with increasing $|\mathbf{r}|$ and then remain nearly constant beyond an intermediate $|\mathbf r|$, thereby exhibiting a saturated plateau. Moreover, the plateau values of $P(\mathbf{r})$ and $\overline{P}(\mathbf{r})$ at the same $U/t$ are quite close, which indicates the predominant interaction contribution to the pair-pair correlation. These features essentially provide clear evidence for the existence of pairing long-range order. In addition, both $P(\mathbf{r})$ and $\overline{P}(\mathbf{r})$ become larger with increasing $U/t$, which is also a characteristic of the interaction-driven BCS-BEC crossover.

\begin{figure}[t]
\centering
\includegraphics[width=1.0\columnwidth]{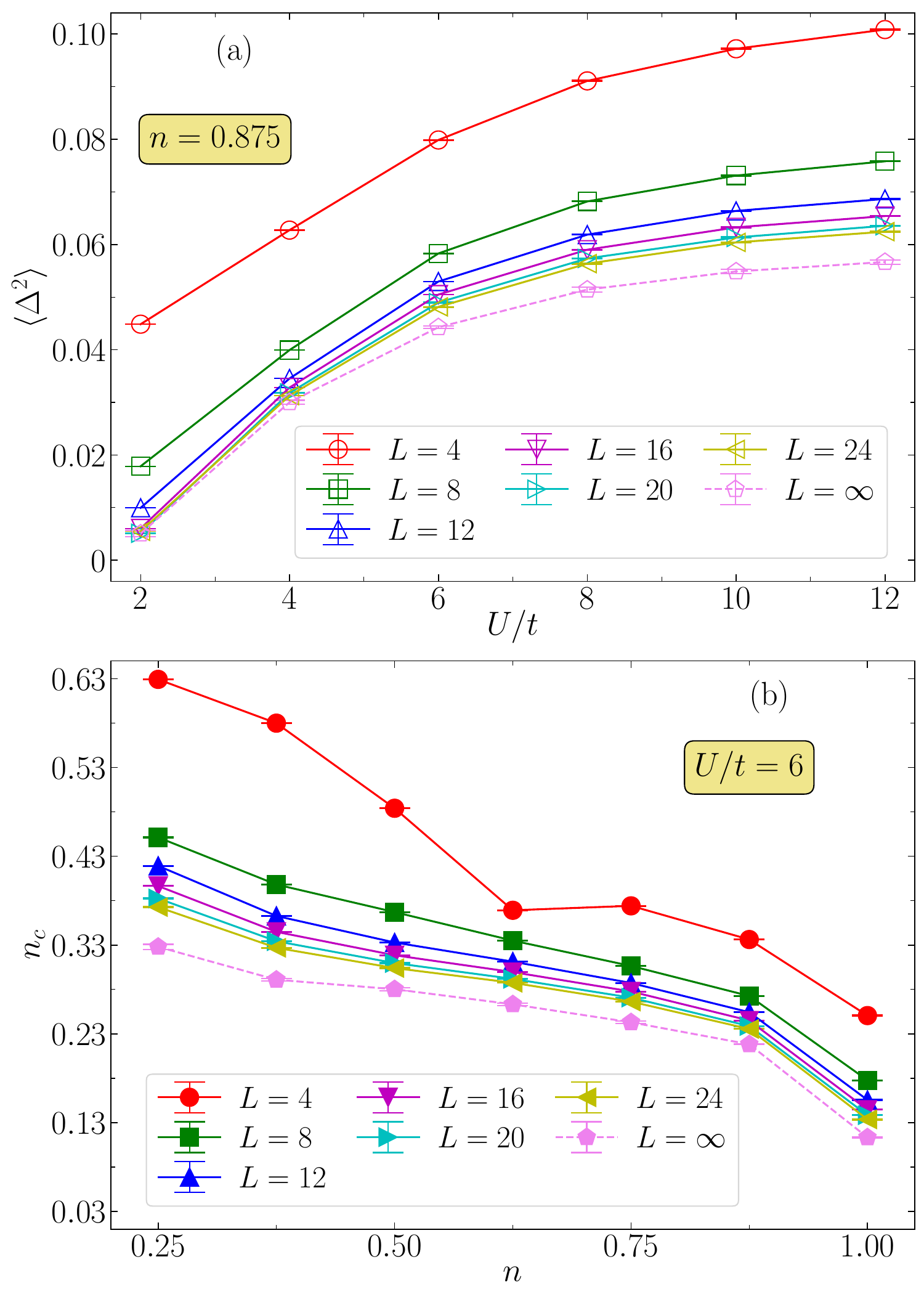}
\caption{
Combined plot of the finite-size (with $L=4\sim 24$ under PBC) and TDL results for mean-squared pairing order $\langle\Delta^2\rangle$ and condensate fraction $n_c$ in the standard 2D attractive Hubbard model~(\ref{eq:HubbModel}). Panel (a) presents $\langle\Delta^2\rangle$ as a function of $U/t$ at fixed $n=0.875$, while panel (b) shows $n_c$ as a function of $n$ at fixed $U/t=6$. In both (a) and (b), the data labeled $L=\infty$ denote the TDL results obtained from finite-size extrapolations similar to those shown in Fig.~\ref{fig:fig06Extrapolate}.
}
\label{fig:fig07finiteLinf}
\end{figure} 

\begin{figure}[t]
\centering
\includegraphics[width=0.977\columnwidth]{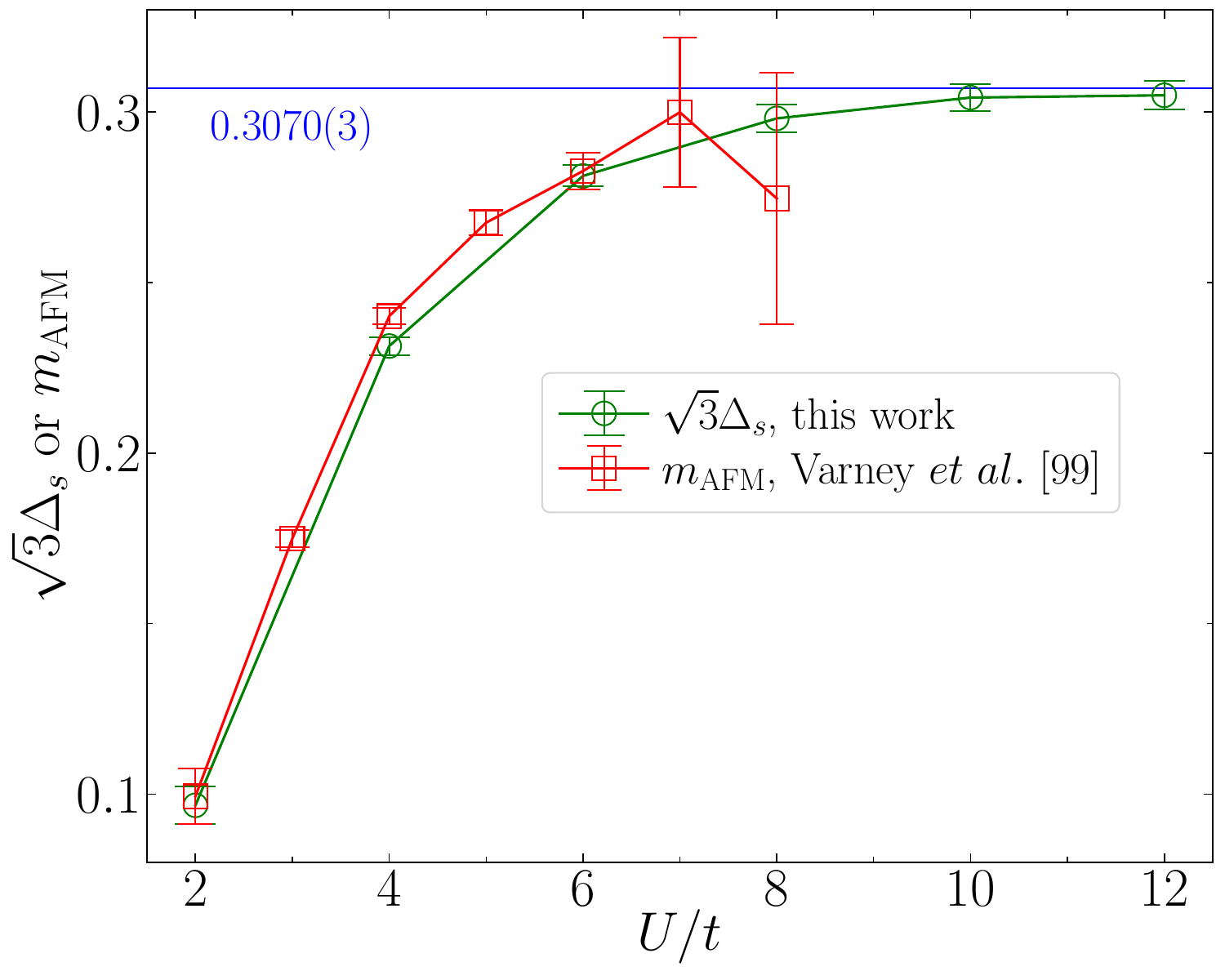}
\caption{
Comparison of the TDL pairing order parameter $\sqrt{3}\Delta_s$ obtained from our AFQMC simulations for the standard attractive Hubbard model at half filling, with the AFM order parameter $m_{\rm AFM}$ reported in Ref.~\cite{Varney2009} for the half-filled repulsive Hubbard model, as a function of $U/t$. The blue solid line marks the strong-coupling Heisenberg-limit value~\cite{Sandvik1997}.
}
\label{fig:fig08Order}
\end{figure}

We next turn to demonstrating the finite-size extrapolations of $n_c$ and $\langle\Delta^2\rangle$ to the TDL, which are shown in Fig.~\ref{fig:fig06Extrapolate} using AFQMC results under PBC. In Fig.~\ref{fig:fig06Extrapolate}(a), we present the finite-size data for $n_c$ and the corresponding extrapolations for three sets of $(U/t,n)$ parameters. For all three cases, the data are nicely described by quadratic polynomial fits in $1/L$, with the corresponding TDL values also included. In Fig.~\ref{fig:fig06Extrapolate}(b), we plot the data of $\langle\Delta^2\rangle$ at $(U/t=6,n=0.625)$, where $P[\mathbf{r}=(L/2,L/2)]$ and $P[\mathbf{r}=(L/2,0)]$ are included as benchmarks to examine the finite-size extrapolation for $\langle\Delta^2\rangle$. As discussed in Sec.~\ref{sec:Observables}, all three quantities indeed produce consistent TDL values within $L\ge 8$ using quadratic polynomial fits in $1/L$. Moreover, we find that both $P[\mathbf{r}=(L/2,L/2)]$ and $P[\mathbf{r}=(L/2,0)]$ exhibit substantially smaller finite-size effects than $\langle\Delta^2\rangle$, which may be attributed to the irrelevant short-range correlations contained in the latter. The finite-size extrapolations of $n_c$ and $\langle\Delta^2\rangle$ using PBC results remain well behaved over a major portion of the parameter regime considered, namely $U/t=2\sim12$ and $n=0.25\sim1.0$, enabling us to obtain precise TDL results. Nevertheless, we find that such extrapolations are challenging at $U/t=2$, where $n_c$ and $\langle\Delta^2\rangle$ results show strong finite-size effects. The same issue was also reported and investigated using calculations under TABC in Ref.~\cite{Karakuzu2018}. To overcome this issue, we incorporate both TABC~\cite{Song2025B,Houmin2026} and finite-size correction~\cite{Zhuotao2025} techniques to ensure the reliability of the TDL results. In Appendix~\ref{sec:FiniteSizeU2}, we show such calculation results at $(U/t=2,n=0.75)$. 

Based on the extrapolated results, we show in Fig.~\ref{fig:fig07finiteLinf} a combined plot of the finite-size and TDL data, with $\langle\Delta^2\rangle$ plotted as a function of $U/t$ at $n=0.875$ in panel (a) and $n_c$ plotted as a function of $n$ at $U/t=6$ in panel (b), to illustrate the finite-size effects. We find that both $\langle\Delta^2\rangle$ and $n_c$ decrease with increasing $L$, and that the finite-size curves with $L\ge8$ follow the same overall trends as the corresponding TDL curve. As $U/t$ increases, $\langle\Delta^2\rangle$ at $n=0.875$ grows monotonically and almost approaches saturation around $U/t=12$, which is consistent with the interaction-driven BCS-BEC crossover. On the other hand, $n_c$ at $U/t=6$ also increases monotonically (for $L\ge8$) with decreasing $n$, reminiscent of the density-induced BCS-BEC crossover (see Appendix~\ref{sec:U6densityCross}). These comparisons demonstrate that our finite-size extrapolations up to $L=24$ reliably capture the essential behaviors of both quantities.

\begin{figure*}[t]
\centering
\includegraphics[width=0.870\linewidth]{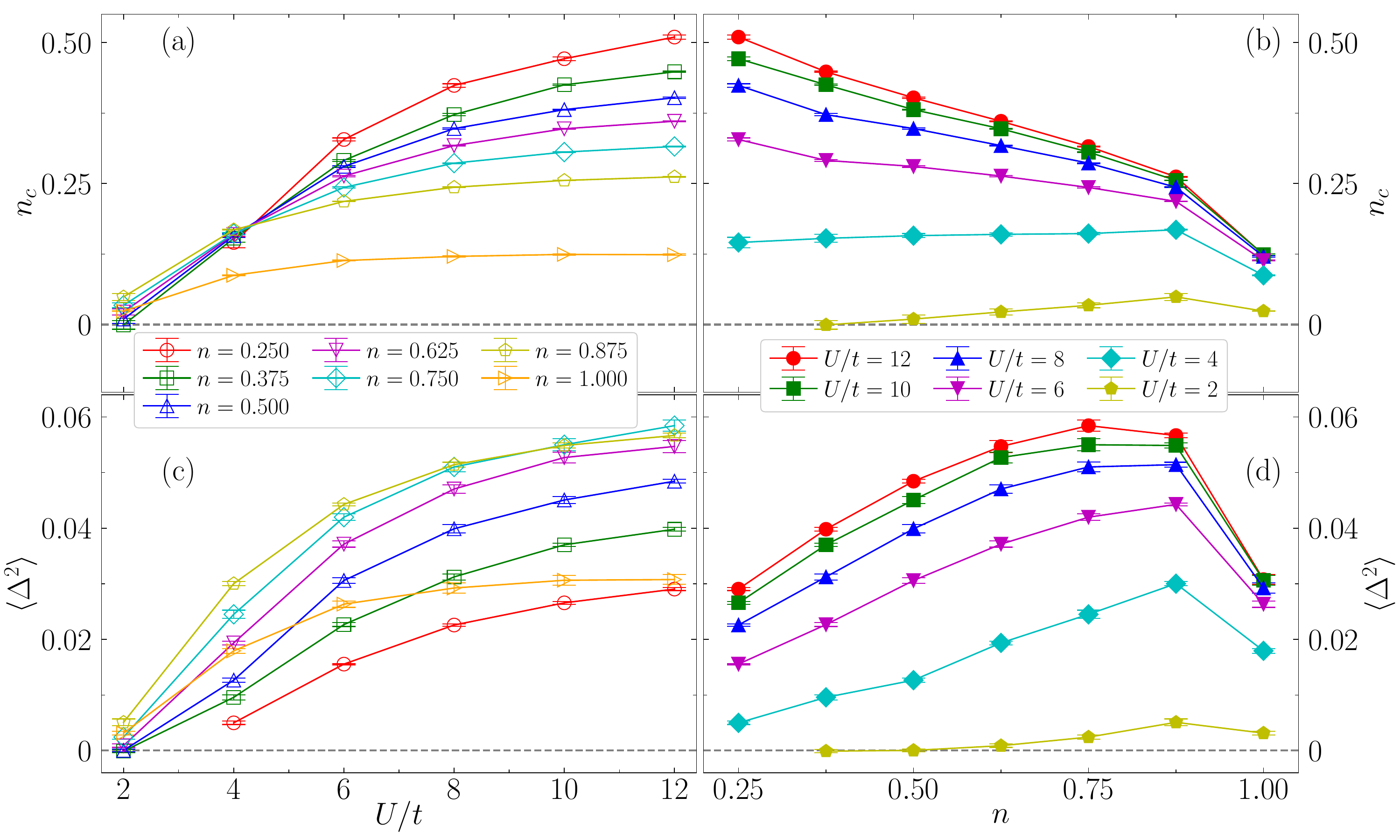}
\caption{
TDL results from finite-size extrapolations for the condensate fraction $n_c$ and mean-squared pairing order $\langle\Delta^2\rangle$ in the standard 2D attractive Hubbard model~(\ref{eq:HubbModel}). Panels (a) and (c) present $n_c$ and $\langle\Delta^2\rangle$, respectively, as functions of $U/t$ for various fermion densities $n=0.25\sim1.0$, while panels (b) and (d) show the corresponding quantities as functions of $n$ for various interaction strengths $U/t=2\sim12$. 
}
\label{fig:fig09TDLorderNc}
\end{figure*}

\begin{figure*}[t]
\centering
\includegraphics[width=0.870\linewidth]{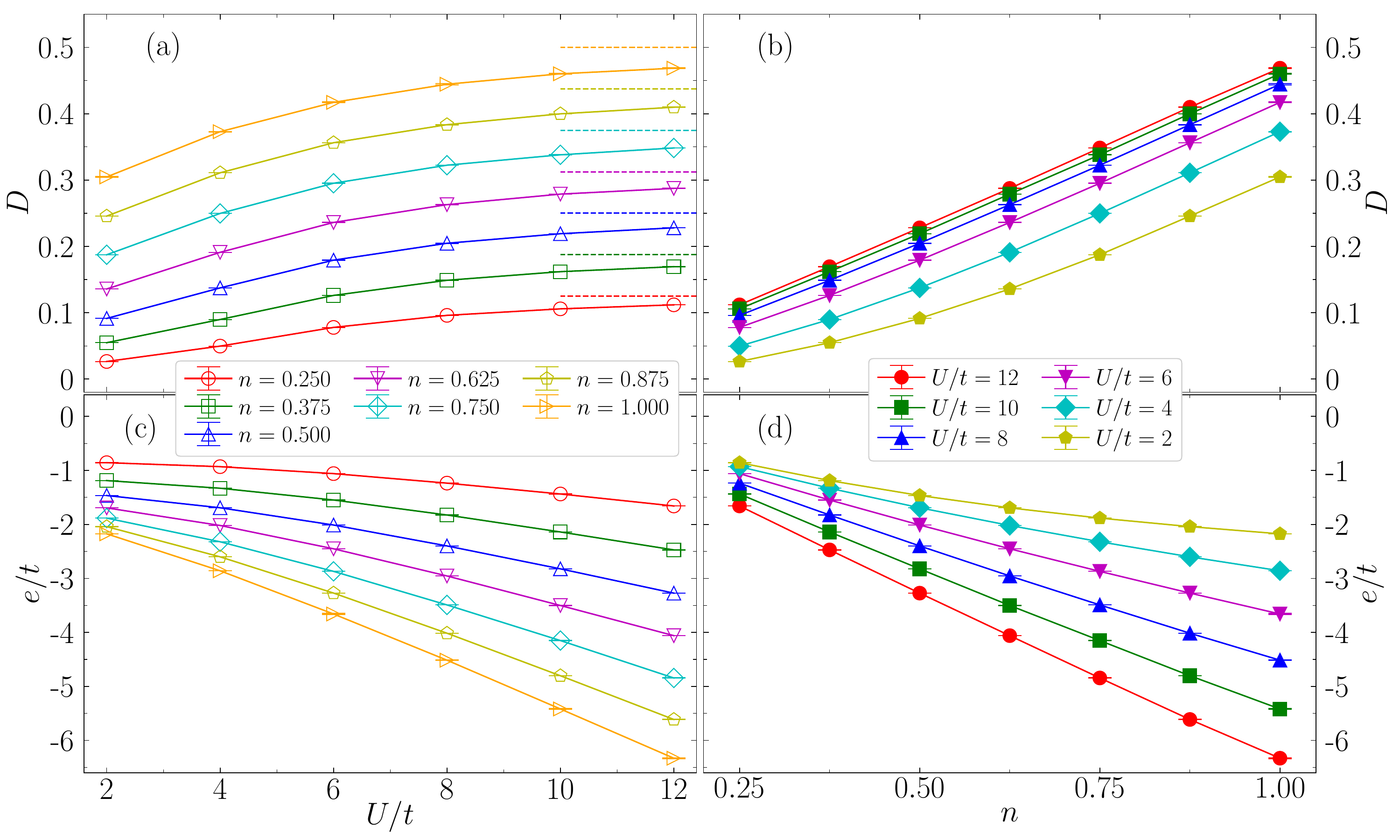}
\caption{
TDL results for the double occupancy $D$ and total energy density $e/t$ in the standard 2D attractive Hubbard model~(\ref{eq:HubbModel}). Panels (a) and (c) present $D$ and $e/t$, respectively, as functions of $U/t$ for various fermion densities $n=0.25\sim1.0$, while panels (b) and (d) show the corresponding quantities as functions of $n$ for various interaction strengths $U/t=2\sim12$. The horizontal dashed lines in (a) indicate the strong-coupling-limit values $D=n/2$ for the data sets with the same colors. The value at $(U/t=2,n=1.0)$ is obtained by finite-size extrapolation of PBC results to the TDL. All other results are obtained with $L=24$, and are verified to converge to the TDL, for which TABC is employed for $U/t=2,n=0.25\sim0.875$, while PBC is used for all other parameter sets. 
}
\label{fig:fig10TDLEngD}
\end{figure*}

At half filling ($n=1$), the pairing order in the attractive Hubbard model can be related to the N\'{e}el antiferromagnetic (AFM) order in the repulsive Hubbard model~\cite{Varney2009} through a partial particle-hole transformation~\cite{Zhuotao2025}. This relation offers  an independent benchmark for our AFQMC calculations. As discussed in Sec.~\ref{sec:TheModel}, the model~(\ref{eq:HubbModel}) at $t_2=0$ and $n=1$ preserves the charge SU(2) symmetry (see Appendix~E in Ref.~\cite{Hille2020} for details). In terms of ${\rm SU(2)}\simeq{\rm SO(3)}\times Z_2$, the SO(3) symmetry rotates the pairing order $\hat{\Delta}_{\mathbf{i}}^+=c_{\mathbf{i}\uparrow}^+c_{\mathbf{i}\downarrow}^+$ and the CDW order $\hat{D}_{\mathbf{i}}=\frac{1}{2}(-1)^{\mathbf{i}}(\hat{n}_{\mathbf{i}\uparrow}+\hat{n}_{\mathbf{i}\downarrow}-1)$ like an O(3) vector as~\cite{Hille2020}
\begin{equation}\begin{aligned}
\label{eq:O3VecCharge}
\hat{\boldsymbol{n}}_{\mathbf{i}} = \big({\rm Re}\hat{\Delta}_{\mathbf{i}}^+,{\rm Im}\hat{\Delta}_{\mathbf{i}}^+,\hat{D}_{\mathbf{i}}\big),
\end{aligned}\end{equation}
with ${\rm Re}\hat{\Delta}_{\mathbf{i}}^+=(\hat{\Delta}_{\mathbf{i}}^+ + \hat{\Delta}_{\mathbf{i}}^{})/2$ and ${\rm Im}\hat{\Delta}_{\mathbf{i}}^+=(\hat{\Delta}_{\mathbf{i}}^+ - \hat{\Delta}_{\mathbf{i}}^{})/(2i)$. Therefore the degeneracy between the pairing order and CDW order at half filling is evident, as numerically verified in both Ref.~\cite{Hille2020} and our calculations. Note that our definition of $P(\mathbf{r})$ in Eq.~(\ref{eq:PairCrFt}) coincides exactly with the correlation functions of ${\rm Re}\hat{\Delta}_{\mathbf{i}}^+$ and ${\rm Im}\hat{\Delta}_{\mathbf{i}}^+$ as
\begin{equation}\begin{aligned}
\label{eq:PairCrFtnew}
P(\mathbf{r}) 
= \frac{1}{N_s}\sum_{\mathbf{i}}\big\langle ({\rm Re}\hat{\Delta}_{\mathbf{i}}^+)({\rm Re}\hat{\Delta}_{\mathbf{i}+\mathbf{r}}^+) \big\rangle,
\end{aligned}\end{equation}
and $\langle({\rm Re}\hat{\Delta}_{\mathbf{i}}^+)({\rm Re}\hat{\Delta}_{\mathbf{i}+\mathbf{r}}^+)\rangle=\langle({\rm Im}\hat{\Delta}_{\mathbf{i}}^+)({\rm Im}\hat{\Delta}_{\mathbf{i}+\mathbf{r}}^+)\rangle$. We then introduce a partial particle-hole transformation $\hat{P}$ acting only on the spin-down operators, defined as~\cite{Zhuotao2025}
\begin{equation}\begin{aligned}
\label{eq:PartialPH}
\hat{P}^+ c_{\mathbf{i}\downarrow}^+\hat{P} &= (-1)^{\mathbf{i}} c_{\mathbf{i}\downarrow} \\
\hat{P}^+ c_{\mathbf{i}\downarrow}\hat{P} &= (-1)^{\mathbf{i}} c_{\mathbf{i}\downarrow}^+,
\end{aligned}\end{equation}
where $(-1)^{\mathbf{i}}$ is the site parity. Under this transformation, the half-filled attractive Hubbard model~(\ref{eq:HubbModel}) is mapped onto the repulsive model, namely $\hat{P}^+\hat{H}(U)\hat{P}=\hat{H}(-U)$. Meanwhile, the vector $\hat{\boldsymbol{n}}_{\mathbf{i}}$ in Eq.~(\ref{eq:O3VecCharge}) is transformed into the spin O(3) vector $\hat{\boldsymbol{s}}_{\mathbf{i}}=(\hat{s}_{\mathbf{i}}^x,\hat{s}_{\mathbf{i}}^y,\hat{s}_{\mathbf{i}}^z)$ [up to a $(-1)^{\mathbf{i}}$ factor accounting for the staggered pattern of the AFM order], which corresponds to the spin SU(2) symmetry. Hence, this transformation establishes a correspondence between the pairing and CDW orders in the half-filled attractive Hubbard model $\hat{H}(U)$ and the AFM orders in spin-$\hat{s}_{\mathbf{i}}^x$, $\hat{s}_{\mathbf{i}}^y$, and $\hat{s}_{\mathbf{i}}^z$ channels of the repulsive Hubbard model $\hat{H}(-U)$. At TDL, the N\'{e}el AFM order parameter $m_{\rm AFM}$ in the repulsive model is defined as~\cite{Varney2009}
\begin{equation}\begin{aligned}
\label{eq:AFMm2}
m_{\rm AFM} = \lim_{L\to\infty}\sqrt{\frac{1}{N_s^2}\sum_{\mathbf{i}\mathbf{j}}(-1)^{\mathbf{i}+\mathbf{j}}\langle\hat{\boldsymbol{s}}_{\mathbf{i}}\cdot\hat{\boldsymbol{s}}_{\mathbf{j}}\rangle}.
\end{aligned}\end{equation}
The ground-state results of $m_{\rm AFM}$ with $U/t$ were obtained via AFQMC in Ref.~\cite{Varney2009}. As $U/t\to\infty$, $m_{\rm AFM}$ should saturate to the value of $0.3070(3)$ in the 2D AFM Heisenberg model~\cite{Sandvik1997}. Since our definition of the pairing order parameter, $\Delta_s=\lim_{L\to\infty}\sqrt{\langle\Delta^2\rangle}$, considers only one of the three degenerate channels in $\hat{\boldsymbol{n}}_{\mathbf{i}}$ [see Eq.~(\ref{eq:O3VecCharge})], the correspondence established above implies $m_{\rm AFM}=\sqrt{3}\Delta_s$. Based on this relation, in Fig.~\ref{fig:fig08Order}, we plot $\sqrt{3}\Delta_s$ as a function of $U/t$ obtained from our AFQMC simulations of the half-filled attractive Hubbard model, and compare it with the $m_{\rm AFM}$ data reported in Ref.~\cite{Varney2009} for the half-filled repulsive Hubbard model. The two groups of data agree within one to two error bars at $U/t=2,4,6$, thus providing a consistency check on our AFQMC simulations. However, the calculations in Ref.~\cite{Varney2009} were limited to $U/t=8$, with the $m_{\rm AFM}$ results at $U/t=7,8$ showing large error bars. In contrast, we extend the AFQMC simulations to $U/t=12$ and achieve high-precision results for the pairing order at TDL. Our data of $\sqrt{3}\Delta_s$ clearly exhibit convergence toward the Heisenberg value as $U/t$ increases, with the result at $U/t=10$ already consistent with this limit within the error bar. Therefore, these results obtained from the attractive Hubbard model can serve as significantly improved estimates for the AFM order parameter in the half-filled repulsive Hubbard model over a broad range of interaction strengths. 

\begin{figure*}[t]
\centering
\includegraphics[width=1.00\linewidth]{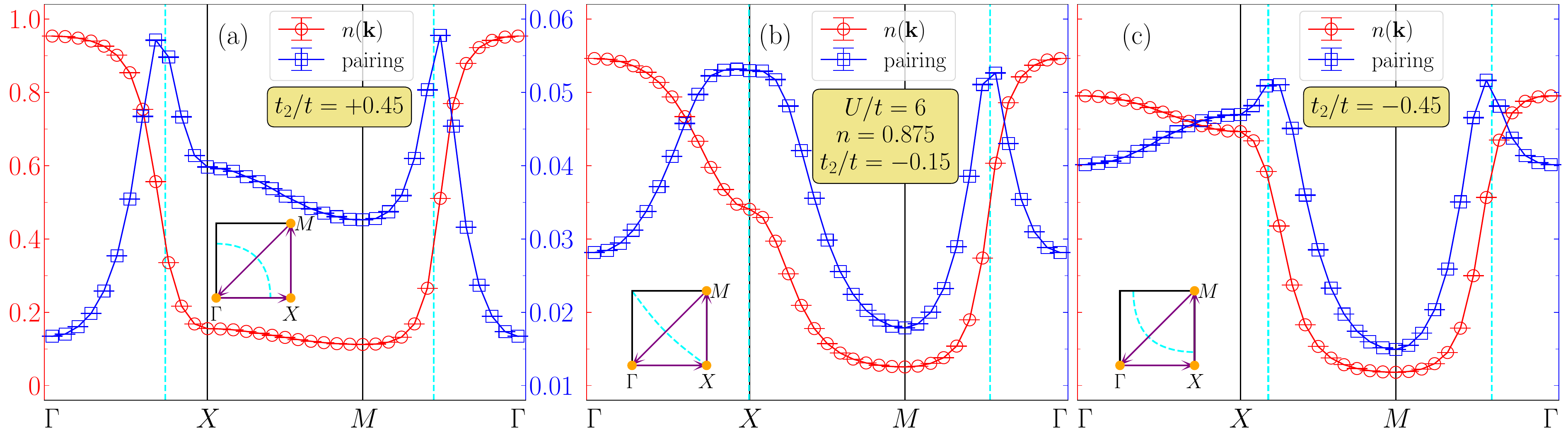}
\caption{
Momentum distribution $n(\mathbf{k})$ and $\mathbf{k}$-space pair wave function $\phi_{\uparrow\downarrow}(\mathbf{k})$ along the high-symmetry path $\Gamma\to X\to M\to\Gamma$ in the Brillouin zone for three NNN hopping amplitudes: (a) $t_2/t=+0.45$, (b) $t_2/t=-0.15$, and (c) $t_2/t=-0.45$. The interaction strength and fermion density are fixed at $U/t=6$ and $n=0.875$, respectively, in all three panels. The left and right $y$-axis scales and tick labels are identical for all panels. Vertical cyan dashed lines indicate the noninteracting Fermi surface, which is shown in the inset of each panel within the quarter Brillouin zone. The calculations are performed with $L=24$ for the 2D attractive Hubbard model~(\ref{eq:HubbModel}).
}
\label{fig:fig11T2PairNk}
\end{figure*}

We then present our main numerical results for $n_c$, $\langle\Delta^2\rangle$, $D$, and $e/t$ in the TDL for the attractive Hubbard model~(\ref{eq:HubbModel}) over the parameter ranges $U/t=2\sim12$ and $n=0.25\sim1.0$. In Fig.~\ref{fig:fig09TDLorderNc}, we plot the data for $n_c$ and $\langle\Delta^2\rangle$ both as functions of $U/t$ at fixed $n$ and as functions of $n$ at fixed $U/t$. As $U/t$ increases, $n_c$ grows monotonically at all densities, which is a typical feature of the BCS-BEC crossover~\cite{Shihao2015,Leggett2012,Mohit2014,Chen2024}. Meanwhile, $\langle\Delta^2\rangle$ also increases, reflecting the continuously enhanced pair-pair correlations, as illustrated in Fig.~\ref{fig:fig05PairCrFt}. For fixed $U/t$, we instead observe interesting nonmonotonic dependences of both quantities on $n$. For $U/t=2$ and $4$, as $n$ decreases, $n_c$ first increases, reaches a broad peak at $n\sim0.875$, and then slowly decreases toward $n\to0$. By contrast, for $U/t\ge6$, the peak behavior disappears, and $n_c$ increases monotonically with decreasing $n$. This behavior is consistent with the evolution toward the BEC regime driven by decreasing density at $U/t\ge6$, as discussed in Sec.~\ref{sec:BCSBEC}. As a comparison, $\langle\Delta^2\rangle$ exhibits a peak behavior versus $n$ for all $U/t$ values, with the peak location shifting from $n\simeq 0.875$ at $U/t=2$ to $n\simeq 0.75$ at $U/t=12$. This nonmonotonic density dependence of $\langle\Delta^2\rangle$ can be explained in terms of two competing effects. At half filling, the pairing and CDW orders are degenerate as a consequence of the charge SU(2) symmetry. Upon doping, this degeneracy is lifted, and the pairing channel is favored, leading to an enhancement of $\langle\Delta^2\rangle$ as $n$ initially decreases. On the small-$n$ side, with further decreasing $n$, the number of fermions available to participate in pairing is progressively reduced, eventually leading to the decrease of $\langle\Delta^2\rangle$. Furthermore, this nonmonotonic behavior of $\langle\Delta^2\rangle$ closely coincides with the density dependence of the superfluid transition temperature reported in Refs.~\cite{Paiva2004,Rodrigo2022}, which also exhibits a peak around $n\sim0.8$ for various $U/t$. Our ground-state results therefore provide a possible physical interpretation of the density dependence of the transition temperature. In addition, the TDL pairing order parameter of the 2D attractive Hubbard model was also presented in Ref.~\cite{Karakuzu2018}, which was limited to $U/t\le4$ and $n=1.0$ and $0.5$. Our results agree well with the reported data. For example, at $U/t=4$, we find $\langle\Delta^2\rangle=0.0179(4)$ at $n=1.0$, and $\langle\Delta^2\rangle=0.0127(4)$ at $n=0.5$, compared with $0.0183(2)$ and $0.0133(3)$, respectively, from Ref.~\cite{Karakuzu2018}. 

In Fig.~\ref{fig:fig10TDLEngD}, we show the TDL results of the double occupancy $D$ and total energy density $e/t$. At all densities, $D$ grows monotonically with increasing $U/t$ and clearly approaches the value of $D=n/2$ at the strong-coupling limit, corresponding to a state in which all fermions form tightly bound on-site pairs. The decreasing $e/t$ with $U/t$ reflects the increasingly dominant contribution of the interaction potential energy (with its density $e_U=-DU$) to the total energy. For all $U/t$, $D$ decreases continuously with decreasing $n$, as expected from the reduced number of fermions available to form on-site pairs. Accordingly, $e/t$ shows monotonic increase with decreasing $n$. Both $D$ and $e/t$ should eventually approach zero in the dilute limit, $n\to0$. Moreover, the kinetic energy density can be readily obtained via $e_K=e-e_U$. These results provide a systematic set of fundamental benchmark data for the 2D attractive Hubbard model, which are easy to be reproduced by various theoretical and numerical methods.

\subsection{The attractive Hubbard model with nonzero \texorpdfstring{$t_2$}{}}
\label{sec:t2Hubbard}

The NNN hopping $t_2$ can modify the superconducting properties of the attractive Hubbard model by altering the single-particle energy dispersion. For $|t_2/t|\le0.5$, the associated Fermi-surface structure changes significantly with $t_2$, but the noninteracting bandwidth remains fixed at $W=8t$. In this subsection, we investigate the evolution of the superconducting properties of the model~(\ref{eq:HubbModel}) by tuning $t_2$ across the range $-0.45\le t_2/t\le0.45$, thus keeping the effective interaction strength $U/W$ constant. We focus on the interaction strength $U/t=6$ and fermion density $n=0.875$. At this density, the local density of states of the noninteracting system features a van Hove singularity (vHs) at $t_2/t\simeq -0.15$, which originates from the saddle points at $X$ points [with $\mathbf{k}_{X}=(\pi,0)$ and $(0,\pi)$] of the dispersion $\varepsilon_{\mathbf{k}}$ in Eq.~(\ref{eq:EkDisp}).

\begin{figure}[t]
\centering
\includegraphics[width=0.97\linewidth]{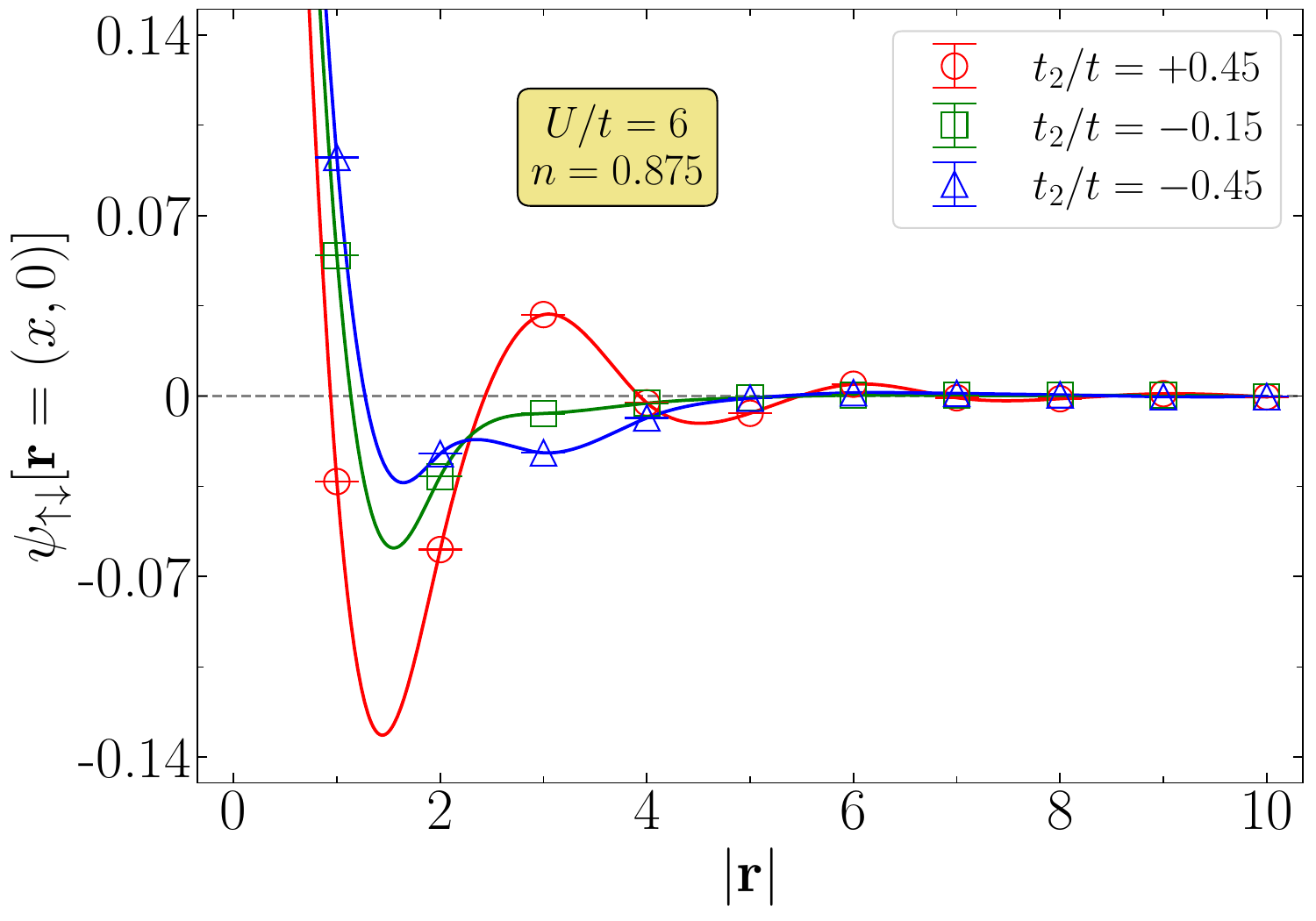}
\caption{
The $\mathbf{r}$-space pair wave function $\psi_{\uparrow\downarrow}(\mathbf{r})$ as a function of $|\mathbf{r}|$ along the $x$ direction [$\mathbf{r}=(x,0)$] with $t_2/t=+0.45$, $-0.15$, $-0.45$. The interaction strength and fermion density are fixed at $U/t=6$ and $n=0.875$, respectively. The solid lines are spline interpolations of the AFQMC data points. These results correspond to those shown in Fig.~\ref{fig:fig11T2PairNk}, and are also obtained with $L=24$.
}
\label{fig:fig12T2RpairWF}
\end{figure}

We first examine the evolution of the fermion pairing with $t_2$. In Fig.~\ref{fig:fig11T2PairNk}, we show the AFQMC results of momentum distribution $n(\mathbf{k})$ and $\mathbf{k}$-space pair wave function $\phi_{\uparrow\downarrow}(\mathbf{k})$ at $t_2/t=+0.45$, $-0.15$, and $-0.45$. These results reveal a BCS-BEC-like crossover behavior as $t_2$ is tuned from positive to negative values. At $t_2/t=+0.45$, $\phi_{\uparrow\downarrow}(\mathbf{k})$ exhibits a pronounced peak near the Fermi surface, indicating the BCS regime. As $t_2/t$ decreases, this peak becomes progressively less prominent, as the relative contrast between the peak and the value at $\Gamma$ point is reduced. This evolution signals a crossover toward a more localized pairing state, although the system has not yet fully reached the BEC regime at $t_2/t=-0.45$. Meanwhile, $n(\mathbf{k})$ also displays signatures of enhanced interaction effects, as its variation across the Fermi surface becomes less sharp and its value at $\Gamma$ point is continuously suppressed as $t_2/t$ decreases. The crossover is also manifested by the corresponding $\mathbf{r}$-space pair wave function $\psi_{\uparrow\downarrow}(\mathbf{r})$. As illustrated in Fig.~\ref{fig:fig12T2RpairWF}, at $t_2/t=+0.45$, $\psi_{\uparrow\downarrow}(\mathbf{r})$ along $\mathbf{r}=(x,0)$ direction clearly shows oscillatory wave structure remaining visible up to $|\mathbf{r}|=8$, indicating BCS-like behavior with spatially extended pairing. As $t_2/t$ is decreased to $-0.15$ and $-0.45$, the wave amplitude of $\psi_{\uparrow\downarrow}(\mathbf{r})$ is suppressed and its curve shrinks to $|\mathbf{r}|\to0$, suggesting increasingly localized pairing.

\begin{figure}[t]
\centering
\includegraphics[width=1.00\linewidth]{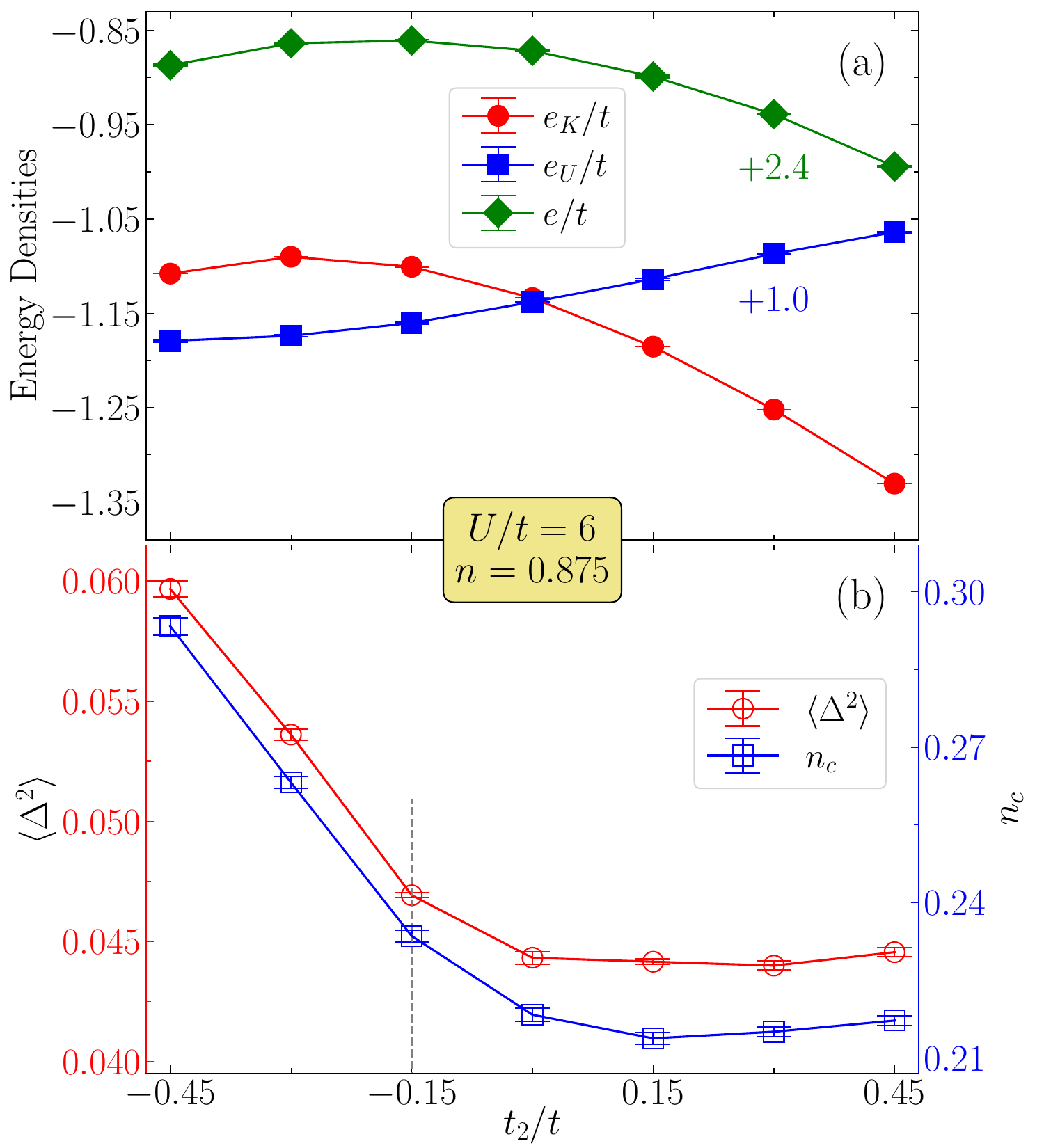}
\caption{
TDL results of (a) the kinetic, interaction, and total energy densities ($e_{K}/t$, $e_{U}/t$, $e/t$), and (b) the mean-squared pairing order $\langle\Delta^2\rangle$ and condensate fraction $n_c$, as functions of the NNN hopping amplitude $t_2/t$ in the 2D attractive Hubbard model~(\ref{eq:HubbModel}). In (a), the data for $e_U/t$ and $e/t$ are shifted by $+1.0$ and $+2.4$, respectively, to fit into the plot. The interaction strength and fermion density are fixed at $U/t=6$ and $n=0.875$, respectively. The gray dashed line at $t_2/t\simeq -0.15$ marks the position where the noninteracting density of states exhibits a van Hove singularity at the Fermi level.
}
\label{fig:fig13T2EngD}
\end{figure}

The BCS-BEC-like crossover as illustrated in Figs.~\ref{fig:fig11T2PairNk} and~\ref{fig:fig12T2RpairWF} and discussed above can be understood from two perspectives. {\it First}, from the energetic picture mentioned in Sec.~\ref{sec:BCSBEC}, decreasing $t_2/t$ from positive to negative corresponds to promoting $\epsilon_b/E_F$ by primarily reducing $E_F$, thereby driving the crossover. As $t_2/t$ varies from $+0.45$ to $-0.45$, the noninteracting Fermi surface evolves from a fermion pocket centered at $\Gamma$ point to a hole pocket centered at $M$ point, as shown in the insets of Fig.~\ref{fig:fig11T2PairNk}. Accordingly, the Fermi energy changes from $E_F/t\simeq6.295$ at $t_2/t=+0.45$, to $\simeq2.800$ at $t_2/t=-0.15$, and $\simeq0.983$ at $t_2/t=-0.45$. {\it Second}, comparing with the $t_2/t\ge0$ case, a negative $t_2/t$ for the NNN hopping raises the kinetic energy (under the condition of fermion density $n\le1$), and thus frustrates nonlocal fermion pairing mediated by successive NN hoppings via increasing its energy cost. Therefore, as $t_2/t$ decreases, particularly into the $t_2/t<0$ regime, the pairing is expected to be more spatially localized, which is consistent with the crossover.

To further explore this $t_2$-induced crossover, we analyze the associated energetics, mean-squared pairing order $\langle\Delta^2\rangle$, and condensate fraction $n_c$. In Fig.~\ref{fig:fig13T2EngD}, we show the TDL results for the kinetic, interaction, and total energy densities (denoted by $e_K/t$, $e_U/t$, and $e/t$), as well as $\langle\Delta^2\rangle$ and $n_c$, as functions of $t_2/t$, with the corresponding finite-size data provided in Ref.~\cite{HubbData2026}. The kinetic energy increases as $t_2/t$ varies from positive to negative, and the interaction energy decreases monotonically which indicates a continuously increasing $D$ as $D=-e_U/U$. Note that the slight decrease in $e_K/t$ from $t_2/t=-0.30$ to $-0.45$ may be related to the specific $U/t=6$, probably lying close to the interaction threshold for the existence of the crossover. The increase in $D$ with decreasing $t_2/t$ indicates enhanced on-site pair formation (as $U/t$ and $n$ are fixed), which is also consistent with the increasingly localized pairing. These observations are in good agreement with our interpretation of the crossover discussed above. Regarding $\langle\Delta^2\rangle$ and $n_c$, both quantities vary only slightly over the positive range of $0\le t_2/t\le +0.45$, and then increase rapidly as $t_2/t$ decreases into the negative range. The latter behavior conforms with the increase of $D$, and supports the evolution toward the BEC regime. Our $\langle\Delta^2\rangle$ results in the range $-0.45\le t_2/t\le 0$ also show a trend similar to that of the superfluid transition temperature reported in Ref.~\cite{Fontenele2026} for the same model~(\ref{eq:HubbModel}) at comparable values of $U/t$ and $n$, thus offering a possible explanation for the latter behavior. Moreover, our numerical results, in particular $\langle\Delta^2\rangle$ and $n_c$, do not exhibit any prominent feature near $t_2/t=-0.15$ [marked by the vertical gray dashed line in Fig.~\ref{fig:fig13T2EngD}(b)], where a vHs appears in the noninteracting system. Theoretically, the vHs with a diverging density of states is generally expected to enhance the associated ordering at weak coupling~\cite{Fontenele2026,Micnas1990}. Such an enhancement of the pairing order is absent in our results, likely because the interaction strength ($U/t=6$) considered here falls outside the weak-coupling regime. 

The $t_2$-induced BCS-BEC-like crossover should persist for other values of $U/t$ and $n$, which demands a more systematic study. We expect the crossover signatures in $n(\mathbf{k})$, $\phi_{\uparrow\downarrow}(\mathbf{k})$, and $\psi_{\uparrow\downarrow}(\mathbf{r})$ to become more pronounced at larger $U/t$, similar to those observed for the density-induced BCS-BEC crossover in Sec.~\ref{sec:BCSBEC}. As for $n$, under the complete particle-hole transformation, the hole-doped case ($n<1$) with $+t_2/t$ can be mapped onto the particle-doped case ($n>1$) with $-t_2/t$~\cite{Haoxu2024}. The transformation applies $c_{\mathbf{i}\sigma}^{+}\to(-1)^{\mathbf{i}}d_{\mathbf{i}\sigma}^{},c_{\mathbf{i}\sigma}^{}\to(-1)^{\mathbf{i}}d_{\mathbf{i}\sigma}^{+}$ to both spin-up and spin-down operators. It changes the sign of the NNN hopping term in the Hamiltonian and transforms the fermion density from $n$ to $2-n$, thus establishing the correspondence between the two doped regimes. Therefore, our numerical results at $n=0.875$ and $+t_2/t$ also correspond to those at $n=1.125$ and $-t_2/t$, with an additional constant shift for certain quantities, e.g., the double occupancy $D(n=1.125)=D(n=0.875)+0.125$. Meanwhile, the $t_2$-induced BCS-BEC-like crossover also exists at $n>1$, and it can be viewed as the particle-hole counterpart of that at the density $2-n$ $(<1)$, with the corresponding $t_2/t$ instead varying from negative to positive in the former case.

\section{Summary and discussion}
\label{sec:Summary}

In summary, we have carried out a systematic study of the superconducting ground state in the 2D attractive Hubbard model using the AFQMC method. In the standard model, we characterize the interaction-driven and density-induced BCS-BEC crossovers using the pair wave functions and fermion momentum distribution. Across these crossovers, we have presented high-precision finite-size (with $L=4\sim 24$) and TDL results for the on-site spin-singlet pairing order, condensate fraction, as well as double occupancy and total energy density, over a broad and physically relevant parameter regime with the interaction strength $U/t=2\sim 12$ and fermion density $n=0.25\sim 1.0$. In the presence of the NNN hopping ($t_2$), we have identified a distinct $t_2$-induced BCS-BEC-like crossover and explored the evolution of the energetics, pairing order, and condensate fraction with $t_2$ at $U/t=6$ and $n=0.875$.

Our work offers a comprehensive view of how fermion pairing and superconductivity in the attractive Hubbard model evolve with the interaction strength, fermion density, and NNN hopping, and further elucidate how these microscopic parameters drive the system across different pairing regimes. The {\it numerically unbiased} and high-precision results presented in our work also establish an extensive {\it numerical dictionary} for the 2D attractive Hubbard model, covering the pair-pair correlation function, mean-squared pairing order, condensate fraction, double occupancy, and energy densities. These data can serve as valuable benchmarks for various theoretical and numerical approaches~\cite{Karakuzu2018,Shi2023,Yuntian2026} and can also offer useful guidance for interpreting the computed superfluid transition temperatures~\cite{Paiva2004,Rodrigo2022,Wang2023,Rodrigo2024,Yuanyao2026,Fontenele2026}. More broadly, our results provide a foundation for further investigations of superconductivity in more complex systems, such as the attractive Hubbard model with spin-orbit coupling~\cite{Peter2017,Song2024} and on composite lattices~\cite{Lee2009,Erhai2006,Zhang2026,Rubem2026,Zujev2014,Rodrigo2024,Prasad2014,Prasad2022,Prasad2024}.

\begin{acknowledgments}
The authors thank Mingpu Qin for helpful discussions. This work was supported by the National Natural Science Foundation of China (under Grants No. 12674308 and No. 12247103), the Quantum Science and Technology-National Science and Technology Major Project (under Grant No. 2021ZD0301900), and the Youth Innovation Team of Shaanxi Universities.
\end{acknowledgments}

\section*{Data Availability}
The data that support the findings of this article are summarized in the dataset reported in Ref.~\cite{HubbData2026} and are publicly available at \url{https://github.com/Ying-Ning/2D-attractive-Hubbard-data}.

\begin{figure}[h]
\includegraphics[width=1.0\columnwidth]{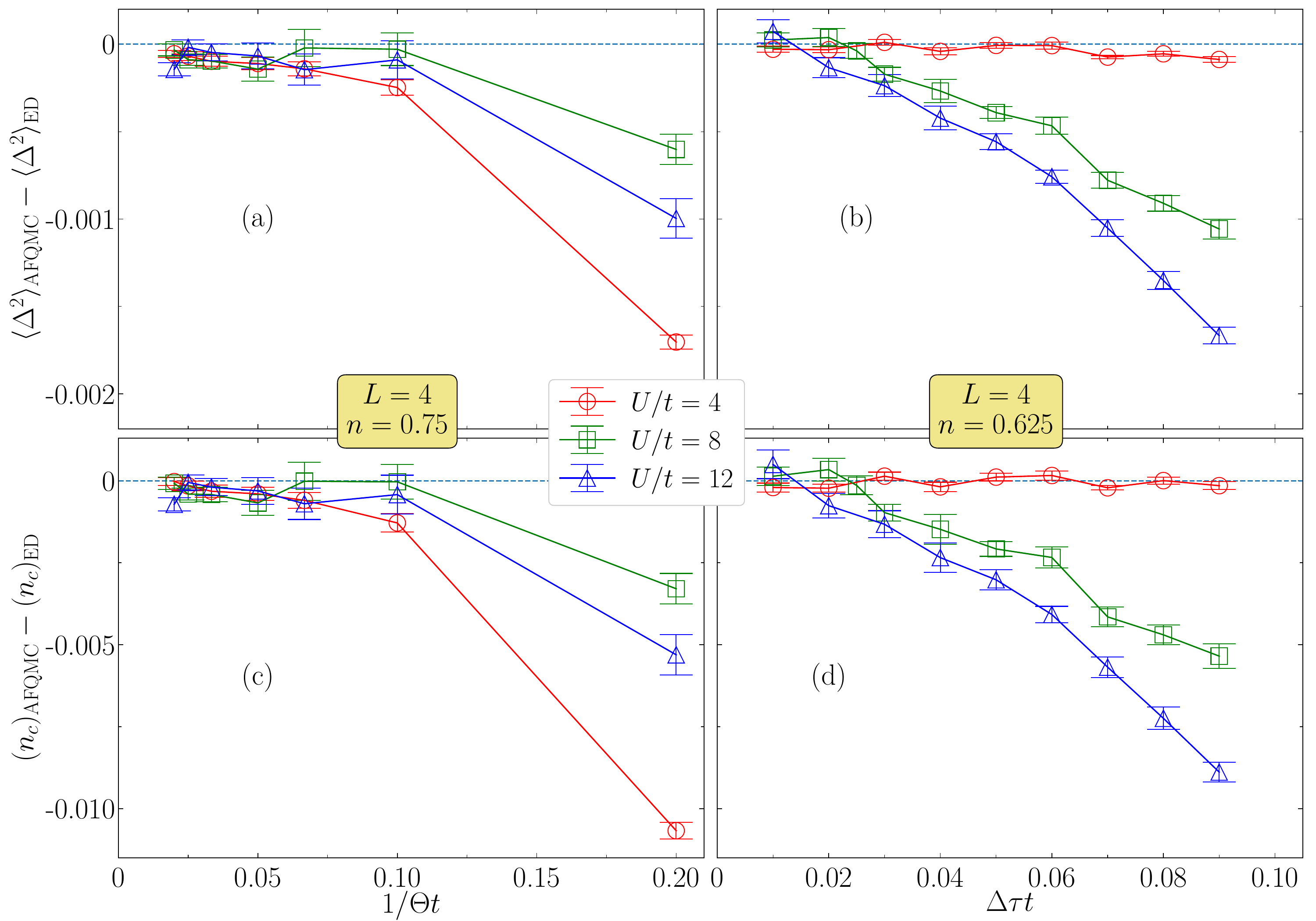}
\caption{
Convergence tests in AFQMC for the projection parameter $\Theta t$ and the imaginary-time step $\Delta\tau t$ on an $L=4$ system. The results are plotted relative to the corresponding ED results. Panels (a) and (c) present the mean-squared pairing order $\langle\Delta^2\rangle$ and condensate fraction $n_c$, respectively, as functions of $1/(\Theta t)$ at $n=0.75$, while panels (b) and (d) show the corresponding quantities as functions of $\Delta\tau t$ at $n=0.625$. All panels include data for $U/t=4$, $8$, and $12$.
}
\label{fig:fig14ConvergeTest}
\end{figure}

\begin{figure*}[t]
\centering
\includegraphics[width=1.00\linewidth]{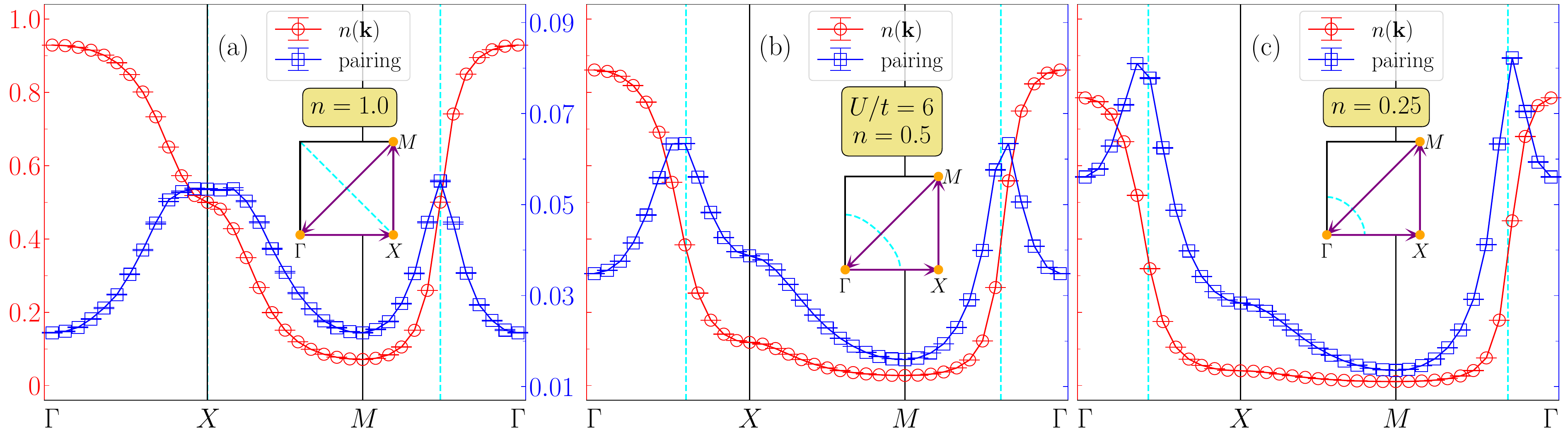}
\caption{
Momentum distribution $n(\mathbf{k})$ and $\mathbf{k}$-space pair wave function $\phi_{\uparrow\downarrow}(\mathbf{k})$ along the high-symmetry path $\Gamma\to X\to M\to\Gamma$ in the Brillouin zone for three fermion densities: (a) $n=1.0$, (b) $n=0.5$, and (c) $n=0.25$. The interaction strength is $U/t=6$ in all three panels. The left and right $y$-axis scales and tick labels are identical for all panels. Vertical cyan dashed lines indicate the noninteracting Fermi surface, which is shown in the inset of each panel within the quarter Brillouin zone. The calculations are performed with $L=24$ for the standard 2D attractive Hubbard model~(\ref{eq:HubbModel}).
}
\label{fig:fig15PairWFnK}
\end{figure*}

\begin{figure}[b]
\centering
\includegraphics[width = 0.99\linewidth]{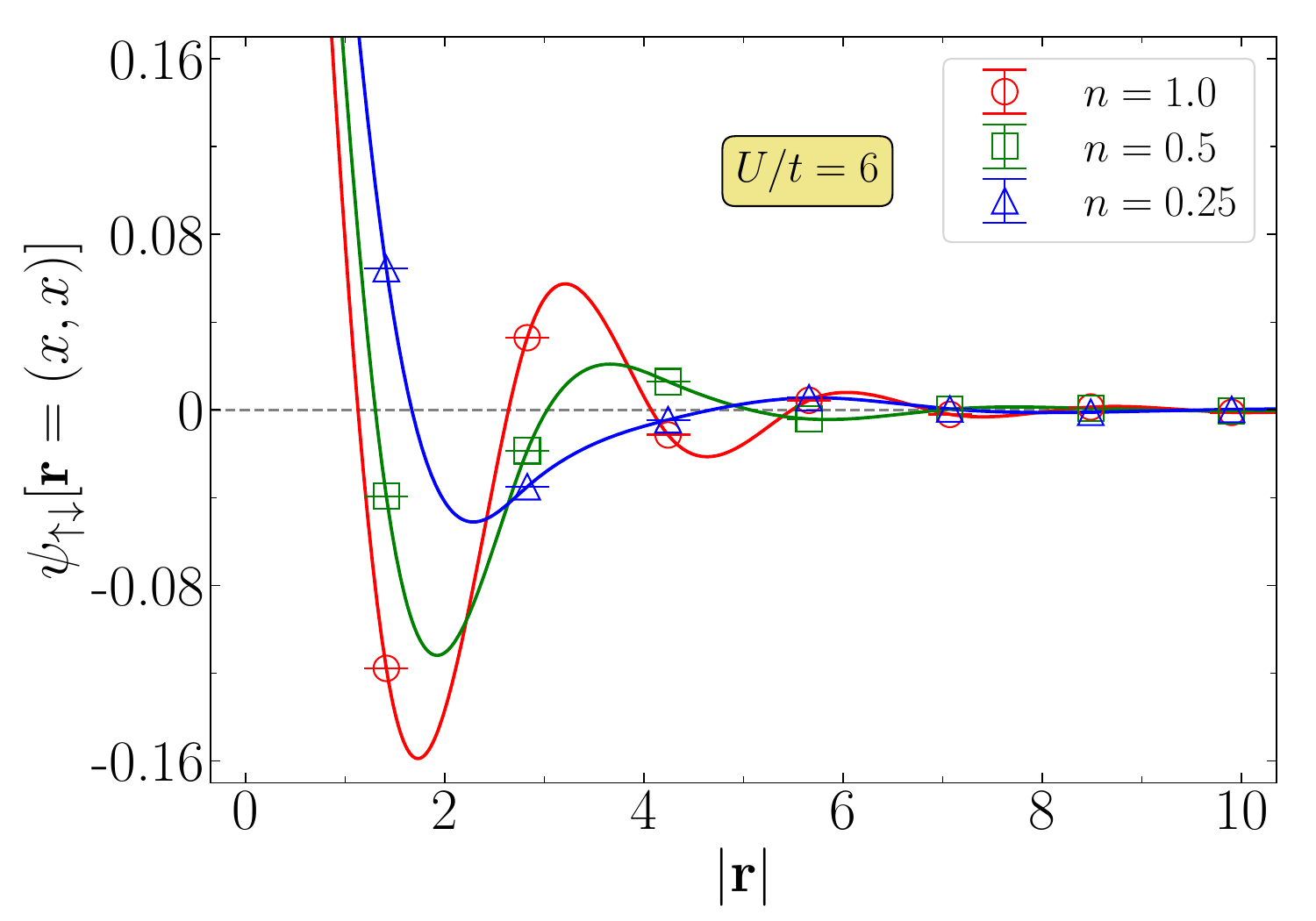}
\caption{
The $\mathbf{r}$-space pair wave function $\psi_{\uparrow\downarrow}(\mathbf{r})$ as a function of $|\mathbf{r}|$ along the $\mathbf{r}=(x,x)$ direction with $n=1.0$, $0.50$, $0.25$. The interaction strength is fixed at $U/t=6$. The solid lines are spline interpolations of the AFQMC data points. These results correspond to those shown in Fig.~\ref{fig:fig15PairWFnK}, and are also obtained with $L=24$.
}
\label{fig:fig16RpairWF}
\end{figure}

\appendix
\section{Convergence tests with \texorpdfstring{$\Theta t$}{} and \texorpdfstring{$\Delta \tau t$}{}}
\label{sec:ConvergeTest}

In Sec.~\ref{sec:AFQMC}, we have briefly introduced the ground-state AFQMC method. To ensure the reliability of the AFQMC simulation, we need to guarantee the convergence to the true many-body ground state of the system, and the elimination of the Trotter error. Accordingly, we have performed the convergence tests on an $L=4$ system for the projection parameter $\Theta t$ and the imaginary-time step $\Delta\tau t$, and have compared the results with ED. Figure~\ref{fig:fig14ConvergeTest} shows the test results. 

In Figs.~\ref{fig:fig14ConvergeTest}(a) and~\ref{fig:fig14ConvergeTest}(c), we plot the AFQMC results of the mean-squared pairing order $\langle\Delta^2\rangle$ and condensate fraction $n_c$, respectively, relative to the corresponding ED results as functions of $1/(\Theta t)$, for the model~(\ref{eq:HubbModel}) with $n=0.75$ at $U/t=4,8,12$. The deviations from zero quantify the projection error. We observe that, at $U/t=8$ and $12$, the AFQMC results converge to the ground-state values more rapidly than at $U/t=4$. This behavior is likely attributed to the enhanced single-particle gap with increasing $U/t$, which typically facilitates a faster convergence to the ground state during the projection process. Based on these results, we find that $\Theta t=30$, $10$, and $10$ are sufficient to achieve convergence for $U/t=4$, $8$, and $12$, respectively. In Fig.~\ref{fig:fig14ConvergeTest}(b) and~\ref{fig:fig14ConvergeTest}(d), we plot similar test results for $\langle\Delta^2\rangle$ and $n_c$ as functions of $\Delta\tau t$ with $n=0.625$ at $U/t=4,8,12$. We observe that the Trotter error is rather weak at $U/t=4$, while it is significant at $U/t=8$ and $12$. This is consistent with the fact that Trotter error in physical observables obtained from AFQMC is proportional to $U$. We determine that $\Delta\tau t = 0.05$, $0.03$, and $0.02$ are small enough to eliminate the Trotter error (the Trotter error is comparable to or smaller than the statistical uncertainties). 

For the AFQMC simulations in Fig.~\ref{fig:fig01QmcED}, we use gradually decreasing $\Theta t$ for increasing $U/t$, ranging from $\Theta t=45$ at $U/t=1$ to $\Theta t=10$ at $U/t=12$ for the $L=4$ system. Meanwhile, we adopt decreasing $\Delta\tau t$ as $U/t$ increases, from $\Delta\tau t=0.05$ at $U/t=1$ to $\Delta\tau t=0.02$ at $U/t=12$, for both $L=4$ and other extended systems. Besides, we use increasing $\Theta t$ as $L$ increases in the simulations, with $\Theta t(L)=\Theta t(L=4)+(L-4)$.

\begin{figure*}[t]
\includegraphics[width = 0.99\linewidth]{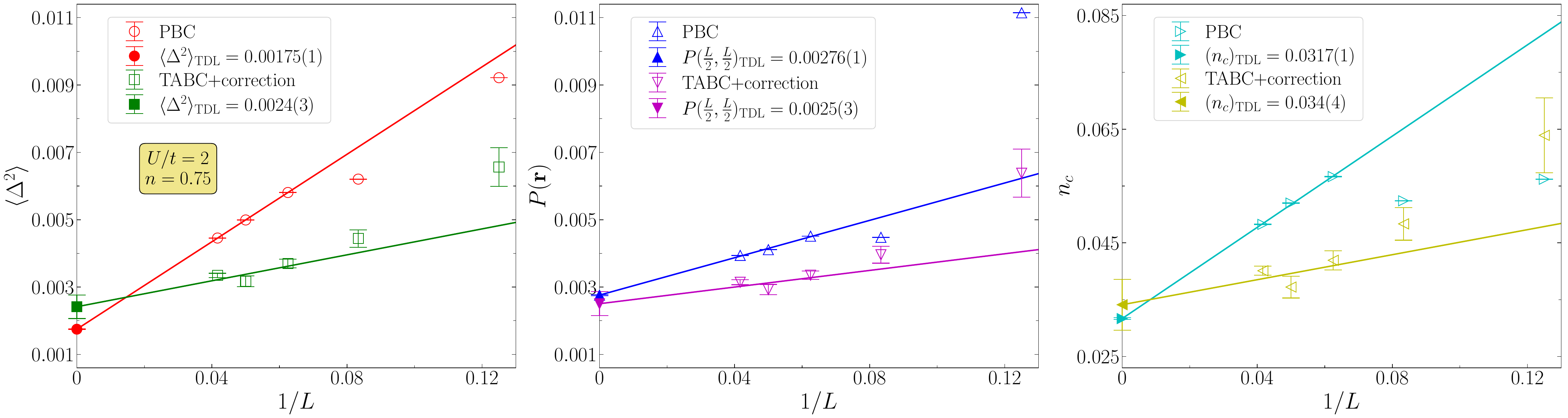}
\caption{Finite-size scaling of (a) the mean-squared pairing order $\langle \hat{\Delta}^2 \rangle$, (b) pair-pair correlation function $P(\mathbf{r})$ at $\mathbf{r}=(L/2,L/2)$, and (c) condensate fraction $n_c$ to the TDL for $U/t=2$, $n=0.75$, using both PBC and TABC. The TABC results are also corrected using the noninteracting model as the reference system. Solid lines are linear fits in $1/L$. The PBC results for $\langle \hat{\Delta}^2 \rangle$ and $P[\mathbf{r}=(L/2,L/2)]$ are inconsistent when extrapolated to $L\to\infty$, whereas the TABC results reach consistent TDL values.
}
\label{fig:fig17extrapolate}
\end{figure*}

\section{Results of \texorpdfstring{$n(\mathbf{k})$}{}, \texorpdfstring{$\phi_{\uparrow\downarrow}(\mathbf{k})$}{}, and \texorpdfstring{$\psi_{\uparrow\downarrow}(\mathbf{r})$}{} for \texorpdfstring{$U/t=6$}{} with varying \texorpdfstring{$n$}{}}
\label{sec:U6densityCross}

In Sec.~\ref{sec:BCSBEC}, we have illustrated the density-induced BCS-BEC crossover at $U/t=12$. Besides, the condensate fraction results in Sec.~\ref{sec:MainResults} begin to exhibit monotonic increase with decreasing $n$ at $U/t=6$, which suggests that the density-induced crossover only exists for $U/t\ge 6$. In this appendix, we show additional results of the fermion momentum distribution $n(\mathbf{k})$ and pair wave functions, $\phi_{\uparrow\downarrow}(\mathbf{k})$ and $\psi_{\uparrow\downarrow}(\mathbf{r})$, for $U/t=6$.

In Fig.~\ref{fig:fig15PairWFnK}, we plot $n(\mathbf{k})$ and $\phi_{\uparrow\downarrow}(\mathbf{k})$ at three fermion densities $n=1.0$, $0.5$, and $0.25$, with the inset showing the corresponding Fermi surface. Although $\phi_{\uparrow\downarrow}(\mathbf{k})$ show a peak near the Fermi surface at all three densities, the relative contrast between the peak and the value at $\Gamma$ point is reduced as $n$ decreases. The broader $y$-axis range of $\phi_{\uparrow\downarrow}(\mathbf{k})$ at lower densities primarily originates from the shrinking Fermi surface and the reduced occupation of fermions with momenta near the corners of the Brillouin zone, similar to that discussed for $U/t=12$. Besides, $n(\mathbf{k})$ show rather smooth variation across the Fermi surface for all the densities, but its value at $\Gamma$ point is continuously suppressed as $n$ decreases. These behaviors indicate the enhanced interaction effects in the system, and thus suggest the crossover behavior with decreasing $n$. Comparing to the results at $U/t=12$ in Fig.~\ref{fig:fig03PairWFnK}, the signatures in $\phi_{\uparrow\downarrow}(\mathbf{k})$ and $n(\mathbf{k})$ at $U/t=6$ are indeed less prominent. In Fig.~\ref{fig:fig16RpairWF}, we illustrate the corresponding results of $\mathbf{r}$-space pair wave function $\psi_{\uparrow\downarrow}(\mathbf{r})$ along $\mathbf{r}=(x,x)$ direction. Similar to those in Fig.~\ref{fig:fig04rPairWvfc}(b), we observe that the oscillatory wave structure is progressively suppressed as $n$ decreases, accompanied by a shrinking spatial extent of $\psi_{\uparrow\downarrow}(\mathbf{r})$ toward $|\mathbf{r}|\to0$. These features further indicate the crossover toward increasingly localized pairing in the system, consistent with the results of $\phi_{\uparrow\downarrow}(\mathbf{k})$ and $n(\mathbf{k})$.

\section{Overcoming the issue associated with finite-size effects for \texorpdfstring{$U/t=2$}{}}
\label{sec:FiniteSizeU2}

In Sec.~\ref{sec:MainResults}, we illustrate and discuss the finite-size extrapolations (see Fig.~\ref{fig:fig06Extrapolate}) to obtain the TDL results. At weak coupling and small fermion density of the model~(\ref{eq:HubbModel}), our AFQMC results show significant finite-size effects, particularly in $\langle\Delta^2\rangle$ and $n_c$. In this appendix, we present additional results to show our data process at $U/t=2$. 

In Fig.~\ref{fig:fig17extrapolate}, we plot the numerical results of $\langle\Delta^2\rangle$, $P[\mathbf{r}=(L/2,L/2)]$, and $n_c$, both under PBC and TABC (with finite-size correction) and their finite-size extrapolations. We observe that the results exhibit oscillations in systems with $L<12$. Moreover, a simple linear fit to the PBC data with $L\ge12$ for $\langle\Delta^2\rangle$ and $P[\mathbf{r}=(L/2,L/2)]$ yields evidently inconsistent TDL results, as included in Figs.~\ref{fig:fig17extrapolate}(a) and~\ref{fig:fig17extrapolate}(b). Therefore, we perform AFQMC simulations under TABC, and apply an additional finite-size correction technique to resolve this issue. The finite-size correction is implemented through~\cite{Zhuotao2025}
\begin{equation}\begin{aligned}
O_{{\rm Corr}}(L) = O(L) - O_{{\rm Ref}}(L) + O_{{\rm Ref}}(\infty),
\end{aligned}\end{equation}
where $O_{\mathrm{Ref}}(L)$ and $O_{\mathrm{Ref}}(\infty)$ denote the values of a specific reference system that can be solved exactly. In this work, we take the noninteracting system as the reference. For each group of the twist angles, the physical observables of the noninteracting system are computed individually, and the average over all twist angles is taken as the reference value $O_{\mathrm{Ref}}(L)$. Under this procedure, the corrected results under TABC of $\langle\Delta^2\rangle$ and $P[\mathbf{r}=(L/2,L/2)]$ extrapolate to consistent TDL values. We perform the same calculation for $n_c$ and also reach reliable TDL data. We implement this combined use of TABC and finite-size correction mainly at $U/t=2$ with $n=0.25\sim 0.875$, mainly for $\langle\Delta^2\rangle$ and $n_c$, and reach the TDL results shown in Fig.~\ref{fig:fig09TDLorderNc}. For the double occupancy and total energy density, TABC typically yields well-converged results up to $L=24$, thus rendering finite-size extrapolation unnecessary.

\bibliography{AttracHubbRef}
\end{document}